\documentclass[letterpaper]{article} 
\usepackage{aaai2026}  
\usepackage{times}  
\usepackage{amsthm}
\usepackage{helvet}  
\usepackage{courier}  
\usepackage[hyphens]{url}  
\usepackage{graphicx} 
\usepackage{natbib}  
\usepackage{caption} 
\usepackage{amssymb}
\usepackage{xcolor}
\usepackage{multirow}
\usepackage{algorithm}
\usepackage{algorithmic}
\usepackage{physics}
\usepackage{booktabs}
\usepackage{tabularx}
\nocopyright
\newcolumntype{Y}{>{\centering\arraybackslash}X}
\usepackage{amsmath}

\newtheorem{theorem}{Theorem}
\usepackage{newfloat}
\usepackage{listings}
\DeclareCaptionStyle{ruled}{labelfont=normalfont,labelsep=colon,strut=off} 
\floatstyle{ruled}
\newfloat{listing}{tb}{lst}{}
\floatname{listing}{Listing}
\title{QuProFS: An Evolutionary Training-free Approach to Efficient Quantum Feature Map Search}

\author {
    Yaswitha Gujju\textsuperscript{\rm 1,\rm 2},
    Romain Harang \textsuperscript{\rm 1},
   Chao Li
   \textsuperscript{\rm 2}, 
  Tetsuo Shibuya
   \textsuperscript{\rm 1},
  Qibin Zhao
   \textsuperscript{\rm 2}
}
\affiliations {
    \textsuperscript{\rm 1}The University of Tokyo, Tokyo, Japan\\
    \textsuperscript{\rm 2}Tensor Learning Team, RIKEN-TRIP, Center for Advanced Intelligence
Project (AIP),
Tokyo, Japan\\
    yaswitha-gujju@g.ecc.u-tokyo.ac.jp

}

\usepackage{bibentry}

\begin{document}

\maketitle

\begin{abstract}
 The quest for effective quantum feature maps for data encoding presents significant challenges, particularly due to the flat training landscapes and lengthy training processes associated with parameterised quantum circuits. To address these issues, we propose an evolutionary training-free quantum architecture search (QAS) framework that employs circuit-based heuristics focused on trainability, hardware robustness, generalisation ability, expressivity, complexity, and kernel-target alignment. By ranking circuit architectures with various proxies, we reduce evaluation costs and incorporate hardware-aware circuits to enhance robustness against noise. We evaluate our approach on classification tasks (using quantum support vector machine) across diverse datasets using both artificial and quantum-generated datasets. Our approach demonstrates competitive accuracy on both simulators and real quantum hardware, surpassing state-of-the-art QAS methods in terms of sampling efficiency and achieving up to a 2× speedup in architecture search runtime.
\end{abstract}


\label{sec:intro}

\section{Introduction}
In the rapidly evolving field of quantum machine learning (QML), quantum kernel methods have emerged as a promising approach for achieving quantum advantage. For instance, quantum kernels have demonstrated exponential learning speedup for the discrete logarithm (DLOG) classification problem~\citep{liu2021rigorous}. While classical kernel methods rely on the kernel trick to implicitly map data into high-dimensional feature spaces, quantum kernels explicitly embed data into Hilbert spaces via quantum feature maps, typically implemented as variational circuits parameterized by the input data—rendering them classically intractable to simulate. Furthermore, the convex optimization landscape of quantum kernel methods offers favorable trainability properties, further motivating their study as a viable route to practical quantum advantage.

Despite these developments, a critical challenge remains: designing efficient and expressive quantum feature maps that faithfully represent real-world data while remaining robust to hardware-induced noise. Existing approaches often face scalability issues in practice~\citep{Wang2021towards}, particularly due to the inefficiency of gradient-based optimization in quantum circuits. In contrast to classical neural networks—where automatic differentiation enables scalable and efficient training—quantum models typically rely on parameter-shift rules for gradient estimation. These can become infeasible as circuit depth and parameter count increase~\citep{abbas2023quantum}.
\begin{figure*}[!t]
    \centering
    \hspace{-2.1cm}
\includegraphics[width=0.8\linewidth]{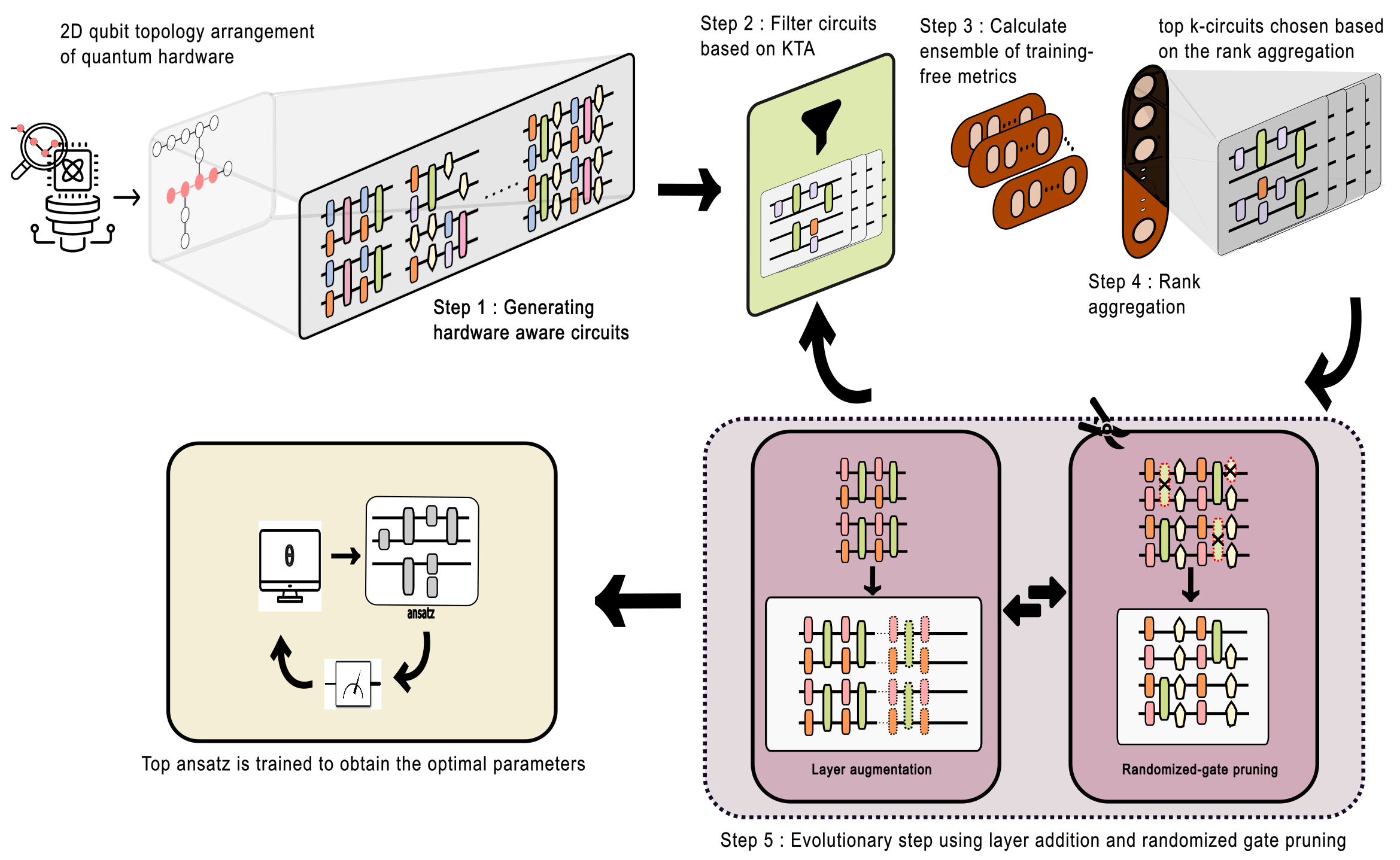}
    \caption{Our QAS pipeline. In the first step, we sample quantum hardware aware circuits as the search space. Following this, we filter 80\% of the circuits using KTA and calculate the proxies for the rest of them in step 2 and 3, respectively. The metrics are ranked appropriately and combined using a novel rank aggregation method in step 4. Top-k circuits are chosen based on the aggregated ranks and are further evolved in step 5 using layer appending and gate pruning. Following new circuit mutations, we go back to step 2 and continue the cycle. }
\label{fig:qka_straight}
\vspace{-0.5cm}
\end{figure*}
Several strategies have been proposed to overcome these limitations. Some approaches (a) introduce trainable parameters into quantum feature maps, while others (b) employ evolutionary or Bayesian algorithms to iteratively refine quantum gate sequences. However, these methods often suffer from high computational costs and limited scalability. Recently, Quantum Architecture Search (QAS) has emerged as a promising paradigm for addressing these challenges by automating the discovery of effective quantum feature maps through meta-model-guided search~\citep{lei2024neural}. Despite its potential, prior work is constrained by the use of fixed ansätze (e.g., the Hardware-Efficient Ansatz~\citep{kandala2017hardware}), which limits circuit diversity and fails to account for hardware constraints. In addition, these approaches often rely on weak evaluation proxies such as expressivity~\citep{he2024training}, increasing the risk of overfitting. Finally, meta-model training typically requires generating new datasets for each task, substantially increasing the computational overhead~\citep{lei2024neural}.

We propose a new architecture, QuProFS (see Figure \ref{fig:qka_straight}), to overcome these key limitations. To broaden the search space, we incorporate both structured and unstructured circuits, integrating hardware properties such as native gate sets, qubit connectivity, and noise characteristics into the search and ranking process to enhance noise robustness.

To overcome the proxy issue, we propose the \emph{first ensemble of training-free metrics} capturing diverse aspects of parameterized quantum circuits (PQC), including expressivity, trainability, hardware robustness, and kernel-data alignment. This lightweight ensemble enables scalable exploration, unlike prior single-proxy approaches. We further design a novel rank aggregation mechanism that combines multiple evaluation criteria to efficiently select quantum circuits. Existing rank aggregation techniques in QAS often prioritize top-performing circuits without sufficiently addressing trade-offs between simulation performance and actual hardware behavior. Our mechanism explicitly incorporates both top-ranked and mid-ranked circuits, allowing a more balanced and robust selection process that better reflects real-hardware performance.

Additionally, many existing studies benchmark on datasets like MNIST and FMNIST, indicating a need for evaluation across a wider variety of datasets, including real-world, artificial, and synthetic ones. Although a performance gap between quantum and classical kernels persists on certain real-world tasks~\citep{schuld2022quantum}, our method significantly narrows this gap while maintaining competitive computational complexity.\vspace{0.4em}\\
\noindent
\textbf{Key contributions of this work:}
\begin{itemize}
    \item We propose the \textbf{first training-free ensemble} of evaluation metrics, repurposed from circuit diagnostics, together with a \textbf{novel rank aggregation} mechanism to enable robust, scalable, and comprehensive quantum circuit selection.
    \item We expand the quantum architecture search space by incorporating structured, unstructured, and hardware-native circuit designs, capturing hardware constraints like qubit connectivity and gate noise.
    \item We benchmark across diverse datasets, achieving higher accuracy with up to \textbf{$2 \times$ speedup} in computational efficiency compared to state-of-the-art QAS methods on both simulator and \textit{real quantum hardware}.
\end{itemize}


\subsection{Related Works}
One line of work focuses on constructing expressive, task-adaptive quantum feature maps by combining data-encoding gates with trainable parameters~\citep{hubregtsen2022training, gentinetta2023quantum, glick2024covariant}, which are optimized to improve performance on specific tasks. However, the reliance on gradient-based methods such as the parameter-shift rule limits scalability as circuit depth and parameter count increase. Another approach employs Bayesian optimization or evolutionary algorithms to search over parameterized quantum circuit structures~\citep{altares2021automatic, pellow2024hybrid, torabian2023compositional}, adapting gate sequences to the data by optimizing circuit topologies based on performance metrics. While more flexible, these methods face high computational overhead due to the combinatorial nature of the search—especially for larger circuits. We combine the structural flexibility of parameterized quantum circuits with the efficiency of training-free evaluation. Rather than optimizing variational parameters or training surrogate models, we propose a fully training-free architecture search strategy. Our method, QuProFS, evaluates candidate circuits using a set of fast, theoretically motivated proxy metrics, enabling scalable and effective circuit selection without the cost of full training.

\subsection{Quantum Architecture Search}

Another line of work focuses on exploring the space of possible circuits efficiently: QAS adapts principles from Neural Architecture Search to parameterized quantum circuits. Early methods used reinforcement learning~\citep{yao2022monte, ostaszewski2021reinforcement} or structured templates with sub-sampling strategies, as seen in QuantumSupernet and QuantumNAS~\citep{du2022quantum, wang2022quantumnas}, to explore circuit spaces efficiently. However, fixed architectures may limit adaptability to diverse data distributions. More recent approaches introduce sparsity~\citep{chen2023quantumsea} or meta-model-based evaluations~\citep{lei2024neural, he2024training, anagolum2024elivagar}, but often suffer from high computational overhead due to surrogate model training and dataset generation. In contrast, our work proposes a fully training-free QAS framework that employs an ensemble of lightweight, theoretically grounded proxy metrics—including expressibility, fidelity, and local effective dimension (LED)—to evaluate circuit suitability without gradient-based optimization or meta-model training, enabling scalable and hardware-aware architecture discovery.

\section{Preliminaries}
In this section, we briefly discuss concepts related to quantum computing to better understand quantum feature maps and their relation quantum kernels. 
\paragraph{Definition and notations.}
Quantum computing generalizes classical bits to qubits, represented as unit vectors in a two-dimensional Hilbert space:
$|\psi\rangle = \alpha |0\rangle + \beta |1\rangle, $ with $|\alpha|^2 + |\beta|^2 = 1.$
Quantum states evolve under unitary operations, implemented via gates such as single-qubit rotations \(R_x(\theta), R_y(\theta), R_z(\theta)\) and two-qubit gates like CNOT. These gates form a universal set capable of expressing any quantum computation. We provide a brief summary of quantum circuit models in Appendix~\ref{annex:qc}.
\paragraph{Quantum Feature Maps.}
Feature maps are used to transform input data \(x \in \mathcal{X}\) into a high-dimensional feature space \(\mathcal{F}\), enabling linear methods to capture nonlinear structure. Classical kernel methods achieve this via implicit inner products by kernel \(\kappa(x, x') = \langle \phi(x), \phi(x') \rangle\) i.e. avoiding explicit computation in \(\mathcal{F}\). Quantum feature maps analogously embed classical data into high-dimensional Hilbert spaces via parameterized quantum circuits. Given an input \(x\), a quantum feature map prepares a quantum state:
\begin{equation}
\ket{\phi(x)} = U(x)\ket{0}^{\otimes n},
\end{equation}
where \(U(x)\) is a unitary encoding circuit. These quantum embeddings serve as the foundation for quantum kernel methods.

\paragraph{Quantum Kernels.}
Quantum kernel methods compute similarity between inputs using overlaps of quantum states. Given data points \(x^i, x^j \in \mathcal{X}\), the quantum kernel is defined as:
\begin{equation} 
\kappa(x^i, x^j) = |\langle \phi(x^i) | \phi(x^j) \rangle|^2.
\label{eq:kernel}
\end{equation}
This overlap is estimated via quantum kernel estimation (QKE), which measures the probability of observing the all-zero bitstring after applying \(U(x^j)^\dagger U(x^i)\) to the initial state~\citep{glick2024covariant}.

While early quantum feature maps encoded only data-dependent parameters, more recent works~\citep{hubregtsen2022training, gentinetta2023quantum, glick2024covariant} introduce additional variational parameters into \(U(x)\) to adapt feature maps to specific datasets. This process, known as \emph{kernel alignment}, improves kernel expressiveness and stability by optimizing the geometry of the feature space to reflect task structure.
\paragraph{Connection to Quantum Neural Networks.}

Quantum neural networks (QNNs) and quantum kernel methods (QKMs) offer distinct but complementary paradigms. QNNs employ parameterized quantum circuits of the form \(V(\boldsymbol{\theta}) U(x) \ket{0}^{\otimes n}\), where \(V(\boldsymbol{\theta})\) is optimized using gradient-based methods. However, such training often encounters barren plateaus—regions of vanishing gradients—posing severe scalability issues in practice. In contrast, QKMs sidestep this challenge by computing quantum kernels from state overlaps and performing classification via convex optimization, without training circuit parameters. This makes them appealing for near-term quantum devices, where circuit depth and gradient quality are constrained.

One work ~\citep{schuld2021supervised} shows that QNNs with fixed measurement observables can be formally reinterpreted as kernel machines, providing a bridge between the two approaches. This insight motivates our use of training-free, kernel-based metrics—such as expressivity, trainability, and alignment—as principled evaluation tools in QAS.

\section{Our Method}
In this section, we introduce our QAS framework called \underline{\textbf{Qu}}antum \underline{\textbf{Pro}}xy-assisted \underline{\textbf{F}}eature \underline{\textbf{S}}earch (QuProFS), leveraging an ensemble of training-free metrics (Figure \ref{tab:proxies_annex}).
 The algorithm operates over a search space \( G \) of parameterized quantum circuits (PQCs), aiming to identify the circuit \( U_{(\Vec{x}, \Theta)} \in G \) that minimizes a loss function \( L \), comparing its output with the ground truth:
\[
\hat{U}^* = \arg \min_{U(\Vec{x}, \Theta) \in G} L\left(U(\Vec{x}, \Theta) |0^N\rangle \right).
\]
An ensemble of proxies capturing distinct circuit properties is computed to be the loss score and rank the different candidate architectures. Below, we detail the steps for search space generation, proxy construction, aggregation, and circuit evolution.
\subsection{Search Space Construction \label{section:search_space}}
The first step in our methodology (Figure~\ref{fig:qka_straight}) is to construct a diverse and hardware-aware search space of quantum circuits. Unlike prior works~\citep{zhang2021neural, he2024training, lei2024neural} that primarily focus on structured circuits based on hardware-efficient ansatz (HEA) templates, our framework integrates both structured and unstructured circuit types to enhance search diversity. This is crucial, as purely random circuit constructions often encounter barren plateaus that prevent effective optimization.
To reduce transpilation overhead, all candidate circuits are composed using the hardware’s native gate set, with two-qubit gates restricted to act only on adjacent qubits according to the device’s connectivity constraints. We further promote diversity by sampling circuits from three distinct categories:
\begin{itemize}
    \item Hardware-Efficient Ansatz (HEA)-style circuits : These circuits are built using a repeating block structure originally proposed by~\citet{kandala2017hardware}. Each block contains single-qubit rotation gates with parameters that can be adjusted, arranged uniformly across the qubits, and these are alternated with two-qubit entangling gates to create correlations between qubits.
\item Covariant Feature Maps : To better capture symmetries inherent in the data, we incorporate covariant quantum feature maps~\citep{glick2024covariant}, which align circuit structure with group-theoretic properties of the input.
\item Noise-aware Unstructured Circuits 
: Leveraging recent noise-based sampling techniques~\citep{anagolum2024elivagar}, we generate circuits that incorporate realistic device noise models, thus reflecting practical hardware conditions.
\end{itemize}

\begin{table*}[t]
  \centering
  \caption{Training-free proxy metrics used to assess quantum circuit quality. See Appendix~\ref{proxies_annex} for formal definitions and derivations.}
  \label{tab:proxies_annex}
  \renewcommand{\arraystretch}{1.4}
  \begin{tabular}{@{}>{\raggedright\arraybackslash}p{4.5cm} >{\arraybackslash}p{11.0cm}@{}}
    \toprule
    \textbf{Metric} & \textbf{Mathematical Expression} \\
    \midrule

    \textbf{Kernel Concentration Indicator} & 
    $\displaystyle \|\mathcal{F} - \mathbf{1}\|_F = \sqrt{\sum_{i=1}^{n} \sum_{j=1}^{n} (f_{ij} - 1)^2}$ \\

    \midrule
    \textbf{Expressivity}~\citep{sim2019expressibility} & 
    $\displaystyle \mathrm{Expr} = \mathrm{KL}\left(\Pr_{\text{emp}} \,\middle\|\, \Pr_{\text{Haar}}\right) = 
    \sum_{j} \Pr_{\text{emp}}(j) \log \left( \frac{\Pr_{\text{emp}}(j)}{\Pr_{\text{Haar}}(j)} \right)$ \\

    \midrule
    \textbf{Kernel Target Alignment (KTA)}~\citep{cristianini2001kernel} & 
    $\displaystyle \mathrm{KTA} = \frac{\langle K, O \rangle_F}{\sqrt{\langle K, K \rangle_F \cdot \langle O, O \rangle_F}}$ \\

    \midrule
    \textbf{Local Effective Dimension}~\citep{abbas2021effective} & 
    $\displaystyle d_{n,\gamma}\left(\mathcal{M}_{\mathcal{B}_\epsilon(\theta^*)}\right) = 
    \frac{2 \log \left( \frac{1}{V_\epsilon} \int_{\mathcal{B}_\epsilon(\theta^*)} 
    \sqrt{ \det \left( I_d + \kappa_{n,\gamma} \bar{F}(\theta) \right)}\, d\theta \right)}
    {\log \kappa_{n,\gamma}}$ \\

    \bottomrule
  \end{tabular}
\end{table*}

\subsection{Circuit Filtering \label{section:filtering}}
Taken together with circuit filtering, we efficiently prune the architecture search space by employing Kernel Target Alignment (KTA)~\citep{cristianini2001kernel}, a computationally lightweight, task-informed proxy for estimating QSVM performance on a representative data subset. KTA measures the alignment between a circuit's kernel matrix and the ideal label kernel via a normalized Frobenius inner product. Candidates with the lowest KTA scores (bottom \( p\% \)) are discarded early, reducing search complexity while retaining promising circuits.

\subsection{Proxies and Rank aggregation}
\label{section:proxies}
To evaluate and rank circuits after filtering, we employ an ensemble of lightweight, training-free proxy metrics (Table~\ref{tab:proxies_annex}). These proxies are designed to capture complementary aspects of circuit quality, including expressivity, trainability, robustness to noise, and alignment with the data distribution. This ensemble approach enables robust, scalable evaluation without requiring full circuit training or execution.\\
\textbf{Kernel Concentration Indicator.} We introduce this simple metric that detects whether a circuit’s kernel is prone to exponential concentration, which can impair class separability in high-dimensional Hilbert spaces.\\
\textbf{Local Effective Dimension.} In order to assess the trainability of each circuit, we estimate their effective dimension using a Monte Carlo approximation of the Fisher Information Matrix, capturing the sensitivity of the model to parameter perturbations.\\
\textbf{Expressivity} We compute the KL divergence between the circuit’s output state distribution and the Haar distribution to assess expressivity, while avoiding overly entangled circuits that may overfit. A full description of each metric, including computational details and theoretical motivation, is provided in Appendix~\ref{proxies_annex}.

\begin{table*}[!th]
\centering
\caption{Accuracy comparison of various methods on artificial quantum datasets.}
\label{table:comparison_artificial}
\begin{tabular}{lccccc}
\toprule
\textbf{Method} & \textbf{Two Curves} & \textbf{Linear Sep.} & \textbf{Hidden Manif.} & \textbf{Bars \& Stripes} & \textbf{Hyperplane} \\
\midrule
\multicolumn{6}{c}{\textbf{Gradient-based QAS}} \\
\midrule
QSEA & 0.80 & 0.93 & 0.87 & 0.78 & 0.87 \\
QNAS & 0.87 & 0.93 & 0.93 & 0.86 & 0.80 \\
\midrule
\multicolumn{6}{c}{\textbf{Training-Free QAS}} \\
\midrule
Elivagar & 0.73 & 0.73 & 0.87 & 0.83 & 0.86 \\
QuProFS (Ours) & \textbf{0.99} & \textbf{0.93} & \textbf{0.89} & \textbf{0.92} & \textbf{0.96} \\
\midrule
\end{tabular}
\end{table*}
\begin{table*}[!h]
\centering
\caption{Accuracy comparison of various methods on classical and quantum datasets.}
\label{table:comparison_2}
\begin{tabular}{lcccc}
\toprule
\textbf{Method} & \textbf{Coset-10} & \textbf{Breast Cancer} & \textbf{MNIST-2} & \textbf{FMNIST-2} \\
\midrule
\multicolumn{5}{c}{\textbf{Gradient-based QAS}} \\
\midrule
QSEA & 0.70 & 0.79 & 0.92 & 0.89 \\
QNAS & 0.70 & 0.78 & 0.95 & 0.89 \\
\midrule
\multicolumn{5}{c}{\textbf{Training-Free QAS}} \\
\midrule
Elivagar & 0.87 & 0.63 & 0.85 & 0.95 \\
QuProFS (Ours) & \textbf{0.86} & \textbf{0.89} & \textbf{0.99} & \textbf{0.97} \\
\midrule
\end{tabular}
\end{table*}
After computing proxy scores, we aggregate rankings using a non-linear rank aggregation method inspired from \citet{lee2024az}, which preserves the relative importance of proxies and reduces outlier influence.
We exclude KTA from this final aggregation since it serves as an initial filter to remove poorly aligned circuits (More details in Appendix).
\begin{figure*}[!t]
    \centering
\includegraphics[width=0.85\linewidth]{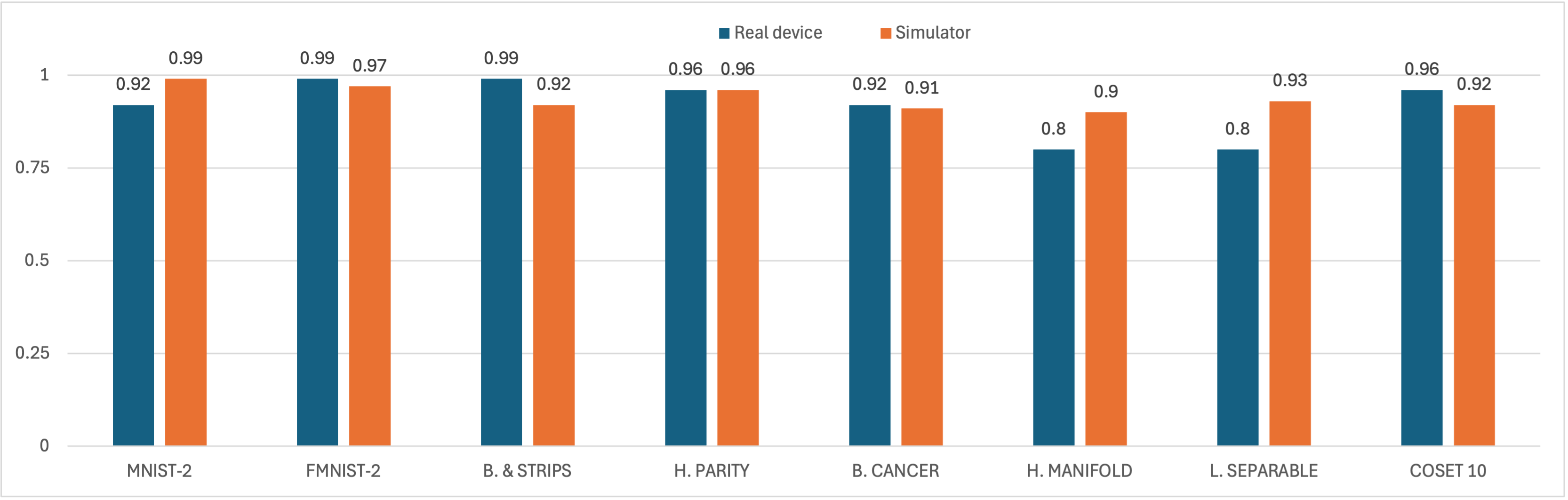}
    \caption{Accuracy results obtained on IBMQ real devices closely match simulator performance, demonstrating the robustness of our approach to hardware noise.}
    \label{fig:real_hw_results}
\end{figure*}
Our main contribution is a modification of the AZ aggregation framework to balance simulator-based performance (with unoptimized variational parameters) and real-hardware robustness. We hypothesize that while proxies like dataset compatibility may reflect performance in simulation, others such as hardware robustness, trainability, and expressivity are more relevant to real-device behavior. To account for these complementary factors, we employ a hybrid aggregation strategy that integrates two distinct ranking schemes within a unified scoring function. To account for these trends, we employ a \textit{hybrid aggregation strategy}  that integrates two distinct ranking schemes within a unified scoring function :
\begin{itemize}
    \item Let the proxies in \(\mathcal{M}_1\) be the \textbf{best-ranked prioritized}, favoring circuits optimal under these criteria and
    \item Proxies in \(\mathcal{M}_2\) are \textbf{mid-ranked prioritized}, preferring circuits that avoid extremes such as excessive depth or expressivity.
\end{itemize}

Given \(Z\) candidate circuits and proxies \(\{P_1, \ldots, P_m\}\), the aggregated score for circuit \(j\) is
\begin{align}
s^{AZ}(j) = \sum_{k \in \mathcal{M}_1} \log\left( \tfrac{r_k^j}{Z} \right) 
+ \sum_{k \in \mathcal{M}_2} \log\left( \tfrac{r_k^j (Z - r_k^j)}{Z} \right)
\end{align}

Following this, the top-\(K\) circuits are selected by maximizing \(s^{AZ}(j)\):
\begin{equation}
\mathcal{T}_K = \operatorname{TopK}_{j \in \{1, \ldots, Z\}} \big( s^{AZ}(j) \big)
\end{equation}
By aggregating these two types of proxy rankings differently, we more accurately reflect the distinct roles each metric plays in circuit evaluation. This method also reduces the influence of outlier circuits that perform exceptionally well on a single metric but poorly on others, leading to a more robust and generalizable architecture selection.

\subsection{Evolutionary operators}
To refine the candidate circuit pool after initial sampling and ranking, we introduce two key operations: \textbf{(a) layer augmentation} and \textbf{(b) randomized gate pruning}. These operations correspond to exploitation and exploration steps, respectively, aimed at improving both circuit performance and search space diversity. \\
\textbf{Layer Augmentation.} In the exploitation phase, we augment each circuit by appending a new layer sampled from the same distribution used in the original search space generation (cf. Section~\ref{section:search_space}Search Space Construction ). This operation exploits existing architectural patterns by deepening circuits that have already shown promise, potentially enhancing their expressivity or trainability without departing from hardware constraints.\\
\textbf{Randomized Gate Pruning.}
To promote diversity and avoid overfitting to early architectural biases, we introduce gate pruning as a complementary exploration mechanism. In each refinement cycle, a subset of circuits undergoes random gate removal at a tunable rate, encouraging exploration while maintaining performance.

\subsection{Runtime Analysis \label{section:theorem}}

Here, we look at the runtime analysis for each iteration of QuProFS. Let us define the key parameters: \(N\) (dataset size), \(M\) ( Subsample size), \(C_{\text{ss}}\) (number of circuit in the search space), \(p_m\) (pruning probability), \(p_f\) (proportion of circuits retained after filtering), \(n\) (number of measurement shots), \(D\) (circuit depth), \(\Theta\) (number of variational parameters \(\theta\) per circuit), and \(G\) (number of gates per circuit). The single forward pass time cost is modeled as $O(nD)$, for the gradient with parameter shift, we model with $O(nD\Theta)$

The total per-iteration time is decomposed as:
\(
T_{\text{iteration}} = T_{\text{evolve}} + T_{\text{metrics}}.
\)
$
T_{\text{metrics}} = C_{\text{ss}} \cdot \Big[
p_f \cdot (M \cdot nD + 2M \cdot nD \cdot \Theta) + M^2 \cdot nD
\Big] \\
= C_{\text{ss}} \cdot nD \cdot \Big[
M \cdot p_f (1 + 2\Theta) + M^2
\Big].
$

\[
T_{\text{evolve}} = C_{\text{ss}} \cdot \left( \Theta D + 0.5 \cdot p_m \cdot G \right) = o(C_{\text{ss}} \cdot MnD).
\]
Hence, the overall per-iteration runtime of QuProFS is
\[
T_{\text{iteration}} = \tilde{O}\left(C_{\text{ss}} \cdot nD \cdot \left[
M (2p_f \Theta + M)
\right]\right).
\]
where the $\tilde{O}$(·) notation suppresses logarithmic factors.

\paragraph{Interpretation.} Theorem~1 shows that the runtime of QuProFS is dominated by proxy evaluation—especially the trainability proxy, which requires costly gradient estimation. To manage this, the dataset is subsampled to a smaller size \( M \ll N \), though the fidelity proxy still incurs a quadratic cost in \( M \). In contrast, the evolutionary step is lightweight, involving simple gate edits with minimal overhead. This separation confirms that QuProFS efficiently allocates computation to evaluating rather than generating circuits, enabling scalable architecture search without full training.

\begin{figure}[!h]
        \centering
        \includegraphics[width=\linewidth]{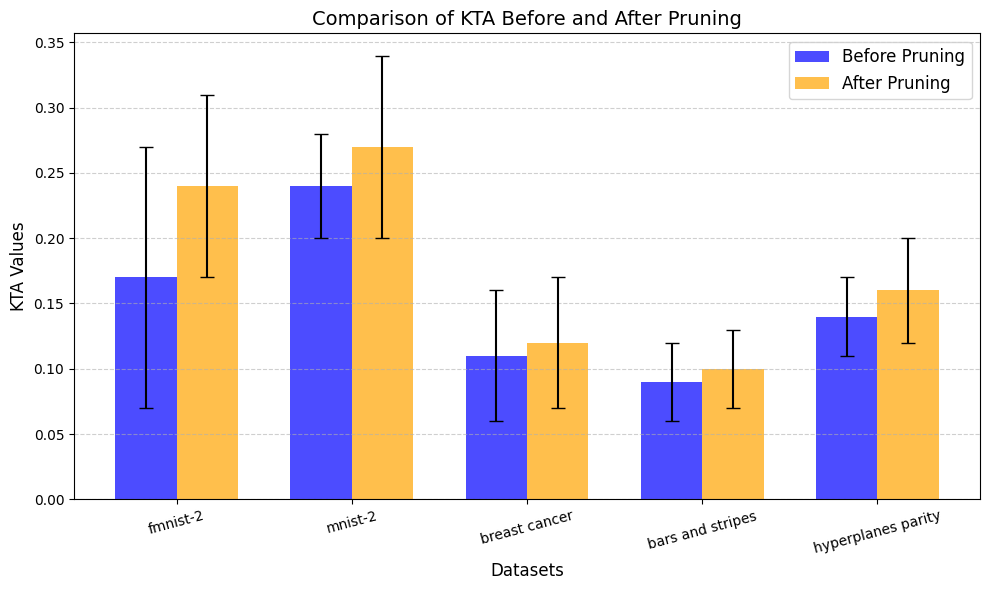}
        \caption{Gate pruning effect with level 0.4 on KTA for a few datasets after 1 iteration.}
        \label{fig:kta_data_pruning}
\end{figure}
\section{Experiments}
We evaluate \textbf{QuProFS} on binary classification tasks spanning classical, synthetic, and quantum-native datasets and compare against state-of-the-art QAS methods. We also assess hardware robustness by executing selected circuits on IBMQ devices. Additional implementation details and extended results are available in the Supplementary Material.

\paragraph{Tasks and Datasets.}
We evaluate QuProFS across a diverse suite of binary classification tasks, categorized into three groups:
\begin{itemize}
    \item \textbf{Classical:} Binary subsets of MNIST~\citep{lecun1998gradient}, FMNIST~\citep{xiao2017fashion}, and the Breast Cancer dataset~\citep{breast_cancer_14}.
    \item \textbf{Quantum-native:} Group-structured datasets such as Coset-10, based on hidden subgroup structure~\citep{glick2024covariant}.
    \item \textbf{Synthetic:} Controlled distributions including Linearly Separable, Hyperplane Parity, Two Curves, and Hidden Manifold datasets, adapted from~\citep{bowles2024betterclassicalsubtleart}.
\end{itemize}

\paragraph{Preprocessing and Experimental Setup.}
We standardize features and apply dimensionality reduction using PCA~\citep{mackiewicz1993principal} to match the desired input size. Datasets are split 80/20 for training/testing. Real-device tests are conducted on 8 representative datasets using selected circuits.

\paragraph{Search Procedure and Hyperparameters.}
Our evolutionary architecture search is run for up to 3 iterations per dataset. For the KTA filtering step, we filter out  \( 80\% \) of the circuits. Full details on evolutionary operators, pruning thresholds, and fitness aggregation can be found in Appendix~\ref{annex:ablation}.

\paragraph{Baselines.}
We benchmark QuProFS against publicly available implementations of leading QAS methods, including QuantumSEA~\citep{chen2023quantumsea}, QuantumNAS~\citep{wang2022quantumnas}, and Elivagar~\citep{anagolum2024elivagar}. 

\begin{figure}[!h]
    \centering
\includegraphics[width=0.8\linewidth]{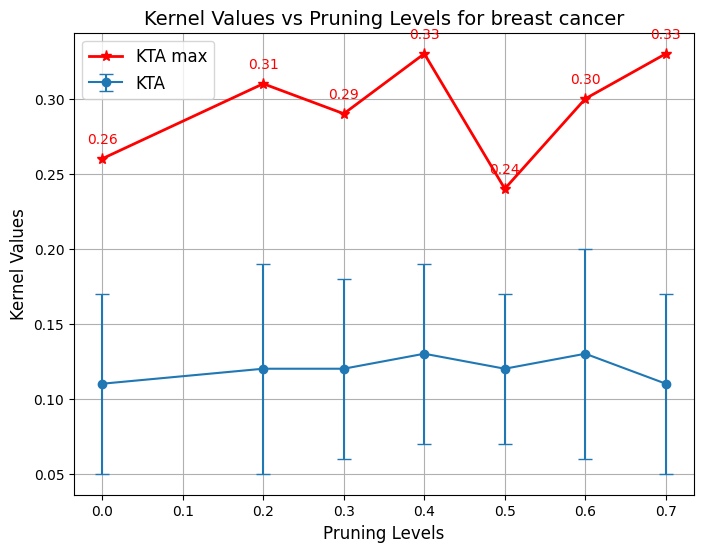}
        \caption{Effect of pruning after 1 iteration on KTA for the Breast Cancer dataset, the blue curve is the average over our top circuits, the blue brackets the 90\% interval and the orange line the highest KTA. The pruning level is the proportion of removed gates.}
        \label{fig:kta_pruning}
        \end{figure}
\vspace{-0.5cm}
\subsection{Results}
We present results in three parts: (i) performance on benchmark datasets, (ii) real quantum hardware evaluation, and (iii) runtime and scaling behavior. QuProFS outperforms existing QAS methods across synthetic, classical, and quantum-native tasks.

\subsubsection{Classification Accuracy on Benchmark Datasets}
QuProFS consistently achieves state-of-the-art accuracy across all dataset types. On synthetic datasets, it outperforms all QAS and training-free baselines, e.g., achieving 0.99 accuracy on Two Curves. On classical benchmarks like MNIST-2 and FMNIST-2, QuProFS reaches 0.99 and 0.97 respectively, surpassing all baselines. On quantum-native datasets, it improves over QSEA and QNAS, demonstrating better handling of structured quantum distributions.
\subsubsection{Performance on Real Quantum Hardware}
We evaluate QuProFS circuits on IBMQ devices to assess robustness under real-device noise and connectivity constraints. Despite the presence of noise and decoherence, QuProFS circuits maintain competitive accuracy, demonstrating robustness and practical viability on real quantum hardware (Figure ~\ref{fig:real_hw_results} and table in appendix~\ref{tab:real_hw_results}) and remain on par with the simulator results. All generated circuits respect the native qubit topology and require no further transpilation.

\subsubsection{Empirical Runtime Comparison}
We set 150 circuits per iteration for two search iterations (three metric evaluations and two evolution rounds). For kernel estimation, we use the full dataset (all $<1500$ samples), while for LED, we subsample to 60 circuits to manage gradient costs, which scale linearly with the number of variational parameters. Runtime is primarily limited by gradient computation, which can be further optimized via parallelization. As shown in Table~\ref{table:run_time_comparison}, QuProFS consistently achieves the lowest runtime across benchmarks, underscoring the efficiency of its training-free, proxy-guided search. 
\begin{table}[!t]
\centering
\caption{Absolute runtime per architecture (in minutes). QuProFS consistently achieves the lowest runtime across datasets.}
\label{table:run_time_comparison}
\resizebox{\columnwidth}{!}{%
\begin{tabular}{lrrrrrr}
\toprule
\textbf{Dataset} & \textbf{QSEA} & \textbf{QNAS} & \textbf{Elivagar} & \textbf{Random} & \textbf{Human} & \textbf{QuProFS (Ours)} \\
\midrule
MNIST-2            & 440  & 865  & 76.6  & 443  & 403  & \textbf{24.3} \\
FMNIST-2           & 439  & 833  & 78.3  & 437  & 403  & \textbf{31.1} \\
Coset              & 205  & 393  & \textbf{26.1}  & 205  & 188  & 36.1 \\
Bars \& Stripes    & 428  & 795  & 48.0  & 424  & 391  & \textbf{26.3} \\
Hyperplane         & 351  & 676  & 45.8  & 336  & 310  & \textbf{22.2} \\
Breast Cancer      & 191  & 410  & 43.0  & 191  & 171  & \textbf{13.2} \\
Two Curves         & 428  & 854  & 43.0  & 428  & 392  & \textbf{14.3} \\
Hidden Manifold    & 343  & 673  & 46.5  & 339  & 311  & \textbf{13.3} \\
Linearly Separable & 428  & 849  & 31.7  & 424  & 391  & \textbf{18.4} \\
\bottomrule
\end{tabular}%
}
\end{table}

 \subsection{Ablation Study 
\label{section:ablation}}
\paragraph{Pruning.} We investigate the impact of circuit pruning through two main approaches: (i) varying the expected proportion of gates pruned—from 20\% to 70\%—and (ii) analyzing changes across various evaluation metrics before and after pruning. Empirically, we observe that this step significantly increases the variance in circuit structure and proxy scores, which is critical for escaping local optima in architecture search.
This refinement strategy—consisting of structured deepening (layer augmentation) and randomized gate pruning —balances architectural intensification and diversification. As we show in Figure~\ref{fig:kta_pruning} and~\ref{fig:kta_data_pruning}, it leads to higher-performing and more robust circuits across simulated and real-device evaluations.

Pruning consistently improves the KTA, with gains extending to downstream tasks like SVM classification, indicating better-aligned feature spaces. It also reduces circuit depth, leading to faster execution and lower noise sensitivity on quantum hardware. Based on these benefits, we combine pruning with layer addition in our mutation strategy to maintain or enhance performance while simplifying circuits for practical deployment.

\section{Discussion \label{section:discussion}}
\paragraph{Conclusion}
The proposed QuProFS algorithm leverages an ensemble of training-free metrics, achieving competitive performance with reduced computational costs. Our results show that this method offers the fastest runtime among QAS benchmarks while maintaining comparable accuracy. By evaluating diverse datasets—real-world, quantum-generated, and artificial—we provide a holistic framework for assessing quantum kernel performance. A key challenge in quantum machine learning remains bridging the gap between classical and quantum kernels. While our work demonstrates that training-free QAS can yield viable solutions without costly circuit training, classical kernels (particularly RBF-based ones) still outperform quantum counterparts on real-world data.

\paragraph{Limitations and Future Work.}
While QuProFS demonstrates promising results, several limitations highlight important directions for future research. The lack of standardized benchmarks in QAS hinders fair comparison and reproducibility due to inconsistent search spaces, metrics, and hardware constraints. While recent work~\citep{lu2023qas} offers public datasets, broader adoption and more diverse, high-dimensional QML benchmarks remain limited. The reliability of training-free proxies remains uncertain, as high proxy scores often fail to predict downstream accuracy, especially across datasets or hardware. This highlights the need for theory linking proxy metrics to generalization, to better guide circuit design. While effective for small- to medium-scale circuits, our evolutionary strategy may face challenges in larger search spaces due to local optima and limited exploration. Scalable performance may require more robust or hybrid search methods. Real-device experiments reveal ongoing challenges. Proxy evaluation is efficient, but training on hardware—especially for kernel methods like QSVM—is slow due to costly kernel estimation. Closing the gap between simulation and hardware is key for practical use. Lastly, to ensure hardware compatibility, we restrict all two-qubit gates to operate on adjacent qubits, respecting native connectivity constraints. This constraint often increases circuit depth and CNOT count, introducing a trade-off between compilation realism and circuit efficiency. A deeper exploration of this trade-off—including alternatives such as SWAP networks or teleportation-based routing—is left for future work.

\section{Acknowledgements}
This work is partly supported by UTokyo Quantum Initiative and the RIKEN TRIP initiative (RIKEN Quantum). We are open to feedback and suggestions, if any. 

\bibliography{main}

\begin{thebibliography}{39}
\providecommand{\natexlab}[1]{#1}

\bibitem[{Abbas et~al.(2023)Abbas, King, Huang, Huggins, Movassagh, Gilboa, and McClean}]{abbas2023quantum}
Abbas, A.; King, R.; Huang, H.-Y.; Huggins, W.~J.; Movassagh, R.; Gilboa, D.; and McClean, J. 2023.
\newblock On quantum backpropagation, information reuse, and cheating measurement collapse.
\newblock \emph{Advances in Neural Information Processing Systems}, 36: 44792--44819.

\bibitem[{Abbas et~al.(2021{\natexlab{a}})Abbas, Sutter, Figalli, and Woerner}]{abbas2021effective}
Abbas, A.; Sutter, D.; Figalli, A.; and Woerner, S. 2021{\natexlab{a}}.
\newblock Effective dimension of machine learning models.
\newblock \emph{arXiv preprint arXiv:2112.04807}.

\bibitem[{Abbas et~al.(2021{\natexlab{b}})Abbas, Sutter, Zoufal, Lucchi, Figalli, and Woerner}]{abbas2021power}
Abbas, A.; Sutter, D.; Zoufal, C.; Lucchi, A.; Figalli, A.; and Woerner, S. 2021{\natexlab{b}}.
\newblock The power of quantum neural networks.
\newblock \emph{Nature Computational Science}, 1(6): 403--409.

\bibitem[{Altares-L{\'o}pez, Ribeiro, and Garc{\'\i}a-Ripoll(2021)}]{altares2021automatic}
Altares-L{\'o}pez, S.; Ribeiro, A.; and Garc{\'\i}a-Ripoll, J.~J. 2021.
\newblock Automatic design of quantum feature maps.
\newblock \emph{Quantum Science and Technology}, 6(4): 045015.

\bibitem[{Anagolum et~al.(2024)Anagolum, Alavisamani, Das, Qureshi, and Shi}]{anagolum2024elivagar}
Anagolum, S.; Alavisamani, N.; Das, P.; Qureshi, M.; and Shi, Y. 2024.
\newblock {\'E}liv{\'a}gar: Efficient quantum circuit search for classification.
\newblock In \emph{Proceedings of the 29th ACM International Conference on Architectural Support for Programming Languages and Operating Systems, Volume 2}, 336--353.

\bibitem[{Bowles, Ahmed, and Schuld(2024{\natexlab{a}})}]{bowles2024betterclassicalsubtleart}
Bowles, J.; Ahmed, S.; and Schuld, M. 2024{\natexlab{a}}.
\newblock Better than classical? The subtle art of benchmarking quantum machine learning models.
\newblock arXiv:2403.07059.

\bibitem[{Bowles, Ahmed, and Schuld(2024{\natexlab{b}})}]{bowles2024better}
Bowles, J.; Ahmed, S.; and Schuld, M. 2024{\natexlab{b}}.
\newblock Better than classical? the subtle art of benchmarking quantum machine learning models.
\newblock \emph{arXiv preprint arXiv:2403.07059}.

\bibitem[{Chen(2024)}]{chen2024evolutionary}
Chen, S. Y.-C. 2024.
\newblock Evolutionary Optimization for Designing Variational Quantum Circuits with High Model Capacity.
\newblock \emph{arXiv preprint arXiv:2412.12484}.

\bibitem[{Chen et~al.(2023)Chen, Zhang, Wang, Gu, Li, Pan, Chong, Han, and Wang}]{chen2023quantumsea}
Chen, T.; Zhang, Z.; Wang, H.; Gu, J.; Li, Z.; Pan, D.~Z.; Chong, F.~T.; Han, S.; and Wang, Z. 2023.
\newblock QuantumSEA: In-Time Sparse Exploration for Noise Adaptive Quantum Circuits.
\newblock In \emph{2023 IEEE International Conference on Quantum Computing and Engineering (QCE)}, volume~1, 51--62. IEEE.

\bibitem[{Cristianini et~al.(2001)Cristianini, Shawe-Taylor, Elisseeff, and Kandola}]{cristianini2001kernel}
Cristianini, N.; Shawe-Taylor, J.; Elisseeff, A.; and Kandola, J. 2001.
\newblock On kernel-target alignment.
\newblock \emph{Advances in neural information processing systems}, 14.

\bibitem[{Du et~al.(2022)Du, Huang, You, Hsieh, and Tao}]{du2022quantum}
Du, Y.; Huang, T.; You, S.; Hsieh, M.-H.; and Tao, D. 2022.
\newblock Quantum circuit architecture search for variational quantum algorithms.
\newblock \emph{npj Quantum Information}, 8(1): 62.

\bibitem[{Garc{\'\i}a-Mart{\'\i}n, Larocca, and Cerezo(2024)}]{garcia2024effects}
Garc{\'\i}a-Mart{\'\i}n, D.; Larocca, M.; and Cerezo, M. 2024.
\newblock Effects of noise on the overparametrization of quantum neural networks.
\newblock \emph{Physical Review Research}, 6(1): 013295.

\bibitem[{Gentinetta et~al.(2023)Gentinetta, Sutter, Zoufal, Fuller, and Woerner}]{gentinetta2023quantum}
Gentinetta, G.; Sutter, D.; Zoufal, C.; Fuller, B.; and Woerner, S. 2023.
\newblock Quantum kernel alignment with stochastic gradient descent.
\newblock In \emph{2023 IEEE International Conference on Quantum Computing and Engineering (QCE)}, volume~1, 256--262. IEEE.

\bibitem[{Glick et~al.(2024)Glick, Gujarati, Corcoles, Kim, Kandala, Gambetta, and Temme}]{glick2024covariant}
Glick, J.~R.; Gujarati, T.~P.; Corcoles, A.~D.; Kim, Y.; Kandala, A.; Gambetta, J.~M.; and Temme, K. 2024.
\newblock Covariant quantum kernels for data with group structure.
\newblock \emph{Nature Physics}, 20(3): 479--483.

\bibitem[{He et~al.(2024)He, Deng, Zheng, Li, and Situ}]{he2024training}
He, Z.; Deng, M.; Zheng, S.; Li, L.; and Situ, H. 2024.
\newblock Training-free quantum architecture search.
\newblock In \emph{Proceedings of the AAAI Conference on Artificial Intelligence}, volume~38, 12430--12438.

\bibitem[{Hubregtsen et~al.(2022)Hubregtsen, Wierichs, Gil-Fuster, Derks, Faehrmann, and Meyer}]{hubregtsen2022training}
Hubregtsen, T.; Wierichs, D.; Gil-Fuster, E.; Derks, P.-J.~H.; Faehrmann, P.~K.; and Meyer, J.~J. 2022.
\newblock Training quantum embedding kernels on near-term quantum computers.
\newblock \emph{Physical Review A}, 106(4): 042431.

\bibitem[{Kandala et~al.(2017)Kandala, Mezzacapo, Temme, Takita, Brink, Chow, and Gambetta}]{kandala2017hardware}
Kandala, A.; Mezzacapo, A.; Temme, K.; Takita, M.; Brink, M.; Chow, J.~M.; and Gambetta, J.~M. 2017.
\newblock Hardware-efficient variational quantum eigensolver for small molecules and quantum magnets.
\newblock \emph{nature}, 549(7671): 242--246.

\bibitem[{LeCun et~al.(1998)LeCun, Bottou, Bengio, and Haffner}]{lecun1998gradient}
LeCun, Y.; Bottou, L.; Bengio, Y.; and Haffner, P. 1998.
\newblock Gradient-based learning applied to document recognition.
\newblock \emph{Proceedings of the IEEE}, 86(11): 2278--2324.

\bibitem[{Lee and Ham(2024)}]{lee2024az}
Lee, J.; and Ham, B. 2024.
\newblock AZ-NAS: Assembling Zero-Cost Proxies for Network Architecture Search.
\newblock In \emph{Proceedings of the IEEE/CVF Conference on Computer Vision and Pattern Recognition}, 5893--5903.

\bibitem[{Lei et~al.(2024)Lei, Du, Mi, Yu, and Liu}]{lei2024neural}
Lei, C.; Du, Y.; Mi, P.; Yu, J.; and Liu, T. 2024.
\newblock Neural auto-designer for enhanced quantum kernels.
\newblock \emph{arXiv preprint arXiv:2401.11098}.

\bibitem[{Liu, Arunachalam, and Temme(2021)}]{liu2021rigorous}
Liu, Y.; Arunachalam, S.; and Temme, K. 2021.
\newblock A rigorous and robust quantum speed-up in supervised machine learning.
\newblock \emph{Nature Physics}, 17(9): 1013--1017.

\bibitem[{Lu et~al.(2023)Lu, Pan, Yan, Shan, Wu, and Yan}]{lu2023qas}
Lu, X.; Pan, K.; Yan, G.; Shan, J.; Wu, W.; and Yan, J. 2023.
\newblock Qas-bench: rethinking quantum architecture search and a benchmark.
\newblock In \emph{International Conference on Machine Learning}, 22880--22898. PMLR.

\bibitem[{Ma{\'c}kiewicz and Ratajczak(1993)}]{mackiewicz1993principal}
Ma{\'c}kiewicz, A.; and Ratajczak, W. 1993.
\newblock Principal components analysis (PCA).
\newblock \emph{Computers \& Geosciences}, 19(3): 303--342.

\bibitem[{Nation and Treinish(2023)}]{mapomatic}
Nation, P.~D.; and Treinish, M. 2023.
\newblock Suppressing Quantum Circuit Errors Due to System Variability.
\newblock \emph{PRX Quantum}, 4: 010327.

\bibitem[{Ostaszewski et~al.(2021)Ostaszewski, Trenkwalder, Masarczyk, Scerri, and Dunjko}]{ostaszewski2021reinforcement}
Ostaszewski, M.; Trenkwalder, L.~M.; Masarczyk, W.; Scerri, E.; and Dunjko, V. 2021.
\newblock Reinforcement learning for optimization of variational quantum circuit architectures.
\newblock \emph{Advances in Neural Information Processing Systems}, 34: 18182--18194.

\bibitem[{Pellow-Jarman et~al.(2024)Pellow-Jarman, Pillay, Sinayskiy, and Petruccione}]{pellow2024hybrid}
Pellow-Jarman, R.; Pillay, A.; Sinayskiy, I.; and Petruccione, F. 2024.
\newblock Hybrid genetic optimization for quantum feature map design.
\newblock \emph{Quantum Machine Intelligence}, 6(2): 45.

\bibitem[{Schuld(2021)}]{schuld2021supervised}
Schuld, M. 2021.
\newblock Supervised quantum machine learning models are kernel methods.
\newblock \emph{arXiv preprint arXiv:2101.11020}.

\bibitem[{Schuld and Killoran(2022)}]{schuld2022quantum}
Schuld, M.; and Killoran, N. 2022.
\newblock Is quantum advantage the right goal for quantum machine learning?
\newblock \emph{Prx Quantum}, 3(3): 030101.

\bibitem[{Sim, Johnson, and Aspuru-Guzik(2019)}]{sim2019expressibility}
Sim, S.; Johnson, P.~D.; and Aspuru-Guzik, A. 2019.
\newblock Expressibility and entangling capability of parameterized quantum circuits for hybrid quantum-classical algorithms.
\newblock \emph{Advanced Quantum Technologies}, 2(12): 1900070.

\bibitem[{Thanasilp et~al.(2024)Thanasilp, Wang, Cerezo, and Holmes}]{thanasilp2024exponential}
Thanasilp, S.; Wang, S.; Cerezo, M.; and Holmes, Z. 2024.
\newblock Exponential concentration in quantum kernel methods.
\newblock \emph{Nature Communications}, 15(1): 5200.

\bibitem[{Torabian and Krems(2023)}]{torabian2023compositional}
Torabian, E.; and Krems, R.~V. 2023.
\newblock Compositional optimization of quantum circuits for quantum kernels of support vector machines.
\newblock \emph{Physical Review Research}, 5(1): 013211.

\bibitem[{Wang et~al.(2022)Wang, Ding, Gu, Lin, Pan, Chong, and Han}]{wang2022quantumnas}
Wang, H.; Ding, Y.; Gu, J.; Lin, Y.; Pan, D.~Z.; Chong, F.~T.; and Han, S. 2022.
\newblock Quantumnas: Noise-adaptive search for robust quantum circuits.
\newblock In \emph{2022 IEEE International Symposium on High-Performance Computer Architecture (HPCA)}, 692--708. IEEE.

\bibitem[{Wang et~al.(2021)Wang, Du, Luo, and Tao}]{Wang2021towards}
Wang, X.; Du, Y.; Luo, Y.; and Tao, D. 2021.
\newblock Towards understanding the power of quantum kernels in the {NISQ} era.
\newblock \emph{{Quantum}}, 5: 531.

\bibitem[{Wolpert(1996)}]{wolpert1996lack}
Wolpert, D.~H. 1996.
\newblock The lack of a priori distinctions between learning algorithms.
\newblock \emph{Neural computation}, 8(7): 1341--1390.

\bibitem[{Wu et~al.(2023)Wu, Yan, Lu, Pan, and Yan}]{wu2023quantumdarts}
Wu, W.; Yan, G.; Lu, X.; Pan, K.; and Yan, J. 2023.
\newblock Quantumdarts: differentiable quantum architecture search for variational quantum algorithms.
\newblock In \emph{International Conference on Machine Learning}, 37745--37764. PMLR.

\bibitem[{Xiao, Rasul, and Vollgraf(2017)}]{xiao2017fashion}
Xiao, H.; Rasul, K.; and Vollgraf, R. 2017.
\newblock Fashion-mnist: a novel image dataset for benchmarking machine learning algorithms.
\newblock \emph{arXiv preprint arXiv:1708.07747}.

\bibitem[{Yao et~al.(2022)Yao, Li, Bukov, Lin, and Ying}]{yao2022monte}
Yao, J.; Li, H.; Bukov, M.; Lin, L.; and Ying, L. 2022.
\newblock Monte carlo tree search based hybrid optimization of variational quantum circuits.
\newblock In \emph{Mathematical and Scientific Machine Learning}, 49--64. PMLR.

\bibitem[{Zhang et~al.(2021)Zhang, Hsieh, Zhang, and Yao}]{zhang2021neural}
Zhang, S.-X.; Hsieh, C.-Y.; Zhang, S.; and Yao, H. 2021.
\newblock Neural predictor based quantum architecture search.
\newblock \emph{Machine Learning: Science and Technology}, 2(4): 045027.

\bibitem[{Zwitter and Soklic(1988)}]{breast_cancer_14}
Zwitter, M.; and Soklic, M. 1988.
\newblock {Breast Cancer}.
\newblock UCI Machine Learning Repository.
\newblock {DOI}: https://doi.org/10.24432/C51P4M.

\end{thebibliography}

\appendix

\clearpage

\section{Appendix}
\subsection{Algorithm}
\begin{algorithm}[H]
\caption{QuProFS: Hardware-Aware Circuit Ranking and Evolution}
\begin{algorithmic}[1]
\STATE \textbf{Input:} Dataset $D$, hardware configuration $H$, number of top circuits $k$, number of iterations $T$
\STATE \textbf{Output:} Top-$k$ circuits after $T$ iterations
\STATE $\mathcal{S}_0 \gets \text{sample\_hardware\_aware\_space}(H)$ \hfill \textcolor{gray}{// Step 1: Initial circuit sampling}
\FOR{$t = 1$ to $T$}
    \STATE $KTA_{\text{sub}} \gets \text{compute\_KTA}(D, \mathcal{S}_{t-1})$ \hfill \textcolor{gray}{// Step 2: Fast KTA-based filtering}
    \STATE $\mathcal{S}_{\text{filtered}} \gets \text{filter\_circuits}(KTA_{\text{sub}}, 0.8)$
    \STATE $M_{\text{metrics}} \gets \text{compute\_proxies}(D, \mathcal{S}_{\text{filtered}})$ \hfill \textcolor{gray}{// Step 3: LED, fidelity, expressibility}
    \STATE $R_{\text{agg}} \gets \text{aggregate\_ranks}(M_{\text{metrics}})$ \hfill \textcolor{gray}{// Step 4: Rank aggregation}
    \STATE $C_{\text{top-}k} \gets \text{select\_top\_k}(R_{\text{agg}}, k)$
    \STATE $\mathcal{S}_t \gets \text{evolve\_circuits}(C_{\text{top-}k})$ \hfill \textcolor{gray}{// Step 5: Mutation via pruning and appending}
\ENDFOR
\STATE \textbf{Return} $C_{\text{top-}k}$
\end{algorithmic}
\end{algorithm}

\subsection{Dataset selection}
The recent work in \citep{bowles2024better} point out the need to choose better datasets to benchmark the quantum models on. This is especially needed and impactful when comparing the performance of the model with other methods.
The no-free-lunch theorems \citep{wolpert1996lack} indicate that for a large set of problems, any model's average performance is equivalent. Therefore, one can only expect good performance on a specific relevant subset of problems, which often results in poorer performance on other subsets. The use of common and infamous datasets like MNIST and Iris have been criticized. Thus, based on comments in \citep{bowles2024better} we also select a list of diverse datasets in addition to those like MNIST.

\begin{table}[ht]
\centering
\caption{Summary of QML benchmarks used in evaluation.}
\label{table:benchmark_summary}
\resizebox{\columnwidth}{!}{%
\begin{tabular}{lccc}
\hline
\textbf{Dataset} & \textbf{Classes} & \textbf{Features} & \textbf{Qubits} \\ \hline
MNIST-2            & 2 & 16 & 4  \\ 
FMNIST-2           & 2 & 16 & 4  \\ 
COSET-10           & 2 & 19 & 10 \\ 
Two Curves         & 2 & 10 & 10 \\ 
Bars \& Stripes    & 2 & 16 & 4  \\ 
Hyperplane Parity  & 2 & 10 & 10 \\ 
Hidden Manifold    & 2 & 10 & 10 \\ 
Linearly Separable & 2 & 10 & 10 \\ 
Breast Cancer      & 2 & 8  & 8  \\ 
\hline
\end{tabular}%
}
\end{table}

\subsection{Kernels}
Quantum kernels typically embed classical data into quantum states through a feature map, subsequently computing their inner products. Typically, quantum kernel feature maps are parameterized on the feature values. A quantum support vector machine (QSVM) uses the obtained quantum kernel and SVM.
The difference between the classical and quantum paradigms comes down to the fact that in classical machine learning, the kernel trick reduces the burden of evaluating the feature map for all pairs of data points, while in the quantum setting, the feature maps are evaluated explicitly for all pairs of points followed by the inner product multiple times based on the number of shots. While kernel methods typically benefit from convex optimization landscapes when provided with an exact Gram matrix, the probabilistic nature of quantum devices poses challenges. The entries of the Gram matrix can only be estimated through repeated measurements, resulting in a statistical approximation rather than an exact matrix. This limitation can negatively impact the performance of quantum kernel methods. As the problem size grows, quantum kernels tend to concentrate exponentially, causing kernel value differences to diminish. Consequently, there is a reliance on the number of shots taken during measurement. It has been observed that with a polynomial shot budget, the model can become overly tuned, less sensitive to the input data, and ultimately struggle to generalize effectively.

\subsection{Quantum Architecture Search}
Under the umbrella of QAS, different appraoches have been studied, mainly deriving inspiration from NAS. The use of reinforcement learning (RL) can be noted in works such as \citep{yao2022monte, ostaszewski2021reinforcement} to perform VQE in order to estimate accurate ground energy estimates. 


These works suffer from architecture evaluation cost due to the large action and state spaces leading to slow convergence. Inspired from NAS, the SuperCircuit-based works such as QuantumSupernet \citep{du2022quantum} and QuantumNAS \citep{wang2022quantumnas} involve optimizing relying on a pre-defined circuit structures to
assist the searching methods making it susceptible finding mainly
inoptimal data embeddings due to embedding-task mismatch. Since optimization on quantum hardware is difficult, these methods suffer from scalability issues due to the high computational runtime for training \citep{anagolum2024elivagar, wu2023quantumdarts}. 

To overcome this limitations, recently
QuantumSEA \citep{chen2023quantumsea} was proposed which instead deals with sparse circuit with vastly lesser number of gates making it feasible for on-chip training and lesser optimization overhead.

Training-free QAS methods have been proposed such as  \citep{he2024training, anagolum2024elivagar}. In both the works, a preliminary metric is used to eliminate a significant portion of the circuits sampled from the search space. 
The filtered circuits then ranked based on metrics like expressivity of the circuit (\citealt{he2024training}).

\begin{table*}[!h]
\centering
\caption{Comparison (accuracy) of various methods on classical and quantum datasets.}
{%
\begin{tabular}{l c c c c}
\hline
\multirow{2}{*}{\textbf{Method}} & \textbf{Coset-10} & \textbf{Breastcancer} & \textbf{MNIST-2} & \textbf{FMNIST-2} \\ \cline{2-5}
 &   Acc. & Acc. & Acc. & Acc.  \\ \hline

\multicolumn{5}{c}{\textbf{QAS}} \\ \hline
\textbf{QSEA} &  0.7 & 0.79 & 0.92 & 0.89  \\ \hline
\textbf{QNAS} &  0.7 & 0.78 & 0.95 & 0.89  \\ \hline

\textbf{Elivagar} & 0.87 & 0.63 & 0.85 & 0.95 \\ \hline
\textbf{QuPROFS(Ours)} &  0.86 $\pm$ 0.01 & 0.92 $\pm$ 0.02 & 0.99 $\pm$ 0.01 & 0.97 $\pm$ 0.01 \\ \hline

\multicolumn{5}{c}{\textbf{Baseline}} \\ \hline
\textbf{Human} &  0.7 & 0.78 & 0.89 & 0.87  \\ \hline
\textbf{Random} &  0.7 & 0.78 & 0.93 & 0.89 \\ \hline
\textbf{qSVM COV} &  0.87 & 0.90 & \textbf{0.98} & \textbf{0.99} \\ \hline
\textbf{qSVM ZZ} &  0.88 & \textbf{0.96} & \textbf{0.99} & \textbf{0.99} \\ \hline
\textbf{qSVM COV optimized} &  0.88 & 0.90 & \textbf{0.99} & \textbf{0.98} \\ \hline
\textbf{qSVM ZZ optimized} &   0.88 & \textbf{0.97} & \textbf{0.98} & \textbf{0.99} \\ \hline

\end{tabular}}
\label{table:comparison_2}
\end{table*}

\begin{table*}[th]
\centering
\caption{Comparison of various methods (accuracy) on artificial datasets using quantum methods.}
\resizebox{\textwidth}{!}{%
\begin{tabular}{l c c c c c}
\hline
\multirow{2}{*}{\textbf{Method}} & \textbf{Two curves} & \textbf{Linearly separable} & \textbf{Hidden manifold} & \textbf{Bars and Stripes} & \textbf{Hyperplane}  \\ \cline{2-6}
 & Acc. & Acc. & Acc. & Acc. & Acc.  \\ \hline

\multicolumn{6}{c}{\textbf{QAS}} \\ \hline
\textbf{QSEA} & 0.80 & 0.93 & 0.87 & 0.78 & 0.87 \\ \hline
\textbf{QNAS} & 0.87 & 0.93 & 0.93 & 0.86 & 0.80  \\ \hline

\textbf{Elivagar} & 0.73 & 0.73 & 0.87 & 0.83 & 0.86  \\ \hline
\textbf{QuPROFS(Ours)} & 0.99 $\pm$ 0.01 & 0.93 $\pm$ 0.01 & 0.89 $\pm$ 0.01 & 0.92 $\pm$ 0.01 & 0.96 $\pm$ 0.01  \\ \hline

\multicolumn{6}{c}{\textbf{Baseline}} \\ \hline
\textbf{Human} & 0.93 & 0.73 & 0.87 & 0.73 & 0.73  \\ \hline
\textbf{Random} & 0.93 & 0.87 & 0.87 & 0.87 & 0.87  \\ \hline
\textbf{qSVM COV} & \textbf{1.00} & 0.82 & 0.92 & 0.95 & \textbf{0.99}  \\ \hline
\textbf{qSVM ZZ} & \textbf{1.00} & 0.82 & \textbf{0.99}  & 0.92 & 0.96 \\ \hline
\textbf{qSVM COV optimized} & \textbf{1.00}  & 0.82 & 0.93 & 0.96 & \textbf{0.99} \\ \hline
\textbf{qSVM ZZ optimized} & \textbf{1.00} & \textbf{1.00} & \textbf{0.99}  & 0.93 & 0.96 \\ \hline
\end{tabular}}
\label{table:comparison}
\end{table*}
\subsection{Rank Aggregation}
\label{sec:rank_aggregation}

After computing all the proxy scores, we aggregate the rankings to select the most promising quantum circuit architectures. A common strategy is to linearly sum the rankings across proxies; however, this approach can obscure poor performance in individual proxies, especially when high-performing and low-performing scores cancel each other out. To address this limitation, we adopt the \textit{non-linear rank aggregation method} proposed in \citep{lee2024az}, which better preserves the relative importance of each proxy and avoids dominance by outliers.

Importantly, we exclude KTA from the final aggregation step since it is used earlier in the pipeline as a coarse filter to discard poorly aligned circuits (cf. Section~\ref{section:filtering}). Since this filtering reduces the diversity of KTA scores, including it in the final ranking aggregation would diminish its discriminative power relative to other proxies.

Our key contribution lies in a principled modification of the AZ aggregation framework ~\ref{sec:rank_aggregation} to explicitly incorporate the trade-offs between performance on simulators (with non-optimized variational parameters) and robustness on real quantum hardware. Unlike linear aggregation, which treats all proxies uniformly and is vulnerable to domination by extreme values, our hybrid strategy allows different aggregation behaviors tailored to the semantics of each proxy class. Specifically, we recognize a divergence: proxies like \textbf{dataset compatibility} are positively correlated with QSVM performance in simulations, while others such as \textbf{hardware robustness}, \textbf{trainability}, and \textbf{expressivity} become more predictive of actual hardware performance.

To model this trade-off, we introduce a \textit{hybrid aggregation strategy} that handles these two classes of proxies differently: 
\begin{itemize}
    \item Proxies in the first group (\(\mathcal{M}_1\)) are \textbf{best-ranked prioritized}, as we want circuits that are maximally effective according to these criteria.
    \item Proxies in the second group (\(\mathcal{M}_2\)) are \textbf{central-ranked prioritized}, reflecting our preference for circuits that avoid extremes (e.g., overly expressive or overly deep circuits).
\end{itemize}

Given \(Z\) candidate circuits and \(m\) proxies \(\{P_1, P_2, \dots, P_m\}\), we compute the aggregated score \(s^{AZ}(j)\) for the \(j\)-th circuit as:
\begin{equation}
\small
s^{AZ}(j) = \sum_{k \in \mathcal{M}_1} \log\left( \frac{\text{Rank}(P_k^j)}{Z} \right) + \sum_{k \in \mathcal{M}_2} \log\left( \frac{\text{Rank}(P_k^j)(Z - \text{Rank}(P_k^j))}{Z} \right)
\end{equation}

The top-\(K\) architectures are then selected according to:
\begin{equation}
\arg\max_{j \in \{1, 2, \dots, Z\},\ j \leq K} \left\{ s^{AZ}(j) \right\}
\end{equation}

This approach guarantees that our ultimate choice achieves an essential balance between optimal simulation efficacy and practical hardware viability, underscoring the robustness of our framework on real quantum hardware.

\subsection{Connection between Quantum Kernels and Quantum Neural Networks}
Quantum neural networks (QNNs) and quantum kernel methods (QKMs) offer two complementary perspectives for quantum machine learning. QNNs are parameterized quantum circuits trained via variational optimization, typically consisting of an encoding circuit $U(x)$ followed by a variational ansatz $V(\boldsymbol{\theta})$ and measurement observable $O$, yielding predictions via $\langle O \rangle_{\boldsymbol{\theta}}$.
\[
\ket{\phi(x, \boldsymbol{\theta})} = V(\boldsymbol{\theta}) U(x) \ket{0}, \quad
\hat{y}(x; \boldsymbol{\theta}) = \langle \phi(x, \boldsymbol{\theta}) | O | \phi(x, \boldsymbol{\theta}) \rangle,
\]

In contrast, QKMs encode classical data $x$ into quantum states $\ket{\phi(x)}$ using a unitary $U(x)$, and define a kernel:
\begin{equation}
\kappa(x, x') = |\langle \phi(x) | \phi(x') \rangle|^2,
\end{equation}
which captures similarity in Hilbert space. This enables classification via convex optimization over training examples, avoiding the non-convexity and barren plateaus that challenge QNN training.

Recent work \citep{schuld2021supervised} shows that QNNs with fixed measurement operators are equivalent to kernel machines. Hence, instead of optimizing circuit parameters, one can directly optimize over a quantum kernel estimated from the overlap of quantum states (quantum kernel estimation). Parameterized feature maps further allow kernel alignment, adapting kernels to specific datasets.

This connection is particularly relevant for our work, where we evaluate circuit robustness and expressivity via kernel-based metrics. In this work, we combine metrics related to the expressibility and trainability of the parameterized quantum circuits for function approximation along with kernel methods to evaluate distances between quantum states, merging flexibility and reliability. Kernel-based training offers favorable convergence, improved noise resilience, and eliminates the need for complex circuit ansätze, making it attractive for near-term quantum devices.
\subsection{Quantum Computing\label{annex:qc}}
Unlike classical computation, where there is the notion of a bit, in quantum computation, the bit is replaced with a qubit defined as $$ |\psi\rangle = \alpha |0\rangle + \beta |1\rangle,$$ where $\alpha$, $\beta$  $\in$ $\mathbb{C}$. The vectors $|0\rangle, |1\rangle$ represent the computational basis in the
two-dimensional Hilbert space $\mathcal{H}$. Furthermore, the absolute squares of the amplitudes $ |\alpha|^2 $ and  $|\beta|^2$ represent the probabilities to measure the qubit in either 0 or 1 state, respectively, and hence satisfy the condition $ |\alpha|^2 + |\beta|^2 = 1$.
Based on this condition, $|\psi\rangle$ as be expressed as 
    $ |\psi\rangle = \cos \frac{\theta}{2} |0\rangle + e^{i\phi}\sin  \frac{\theta}{2} |1\rangle$
    where $0 \leq \theta \leq \pi$  and $0 \leq \phi \leq 2\pi$ are real numbers. To transform the quantum states from one state to another, we use the quantum gates represented using unitary matrices, which can be single-qubit or multi-qubit. Using unitary gates ensures that the condition on the amplitude-based probabilities is maintained even after the transformation. Examples of single qubit non-parameterized quantum gates include $X$ gate,
the $Y$ gate and the $Z$ gate which can be used to manipulate a qubit's basis state, amplitude, or phase, respectively. Another gate called the Hadamard gate can be used to put a
qubit with $\beta = 0 \hspace{1mm}  (\alpha = 0)$ into an equal superposition. $\alpha = \frac{1}{\sqrt{2}}, \beta = {\displaystyle \pm }\frac{1}{\sqrt{2}}$ (the Hadamard
or $H$-gate). Under parameterized single qubit gates, we have 
\( R_x(\theta) \), \( R_y(\theta) \), and \( R_z(\theta) \) are defined as 
$R_x(\theta) = e^{-i \frac{\theta}{2} \sigma_x}$ (similar for $R_y$ and $R_z$). 

In the case of multi-qubit gates, a control and target qubit exists, such that a single qubit
operation is applied to the target qubit only if the control qubit is in a certain state. Most common examples include the (CNOT or CX) gate, a two-qubit controlled-NOT. If the control qubit is in state $|1\rangle$, the CNOT gate flips the basis state of the target qubit. The universal gate set of arbitrary one-qubit rotation gates and two-qubit controlled-NOT (or CNOT) gates can implement any quantum operation using a combination of these basic gates.
\subsection{Circuit Filtering \label{section:filtering}}
In the second step of the proposed algorithm (as illustrated in Figure~\ref{fig:qka_straight}), we introduce a hyperparameter to control the number of filtered circuits. To achieve this, we utilise kernel target alignment (KTA) to estimate the potential accuracy of Quantum Support Vector Machines (QSVM) due to its strong correlation with architectural performance. KTA has also been employed as a loss function for training parameterised feature maps~\citep{hubregtsen2022training}. It measures the similarity between two kernels or the degree of alignment between a kernel and a dataset. In our research, we apply KTA to a subsampled dataset (see \ref{KTA_annex}) to evaluate the compatibility between the task and the selected kernel. Since KTA provides a reliable approximation of circuit performance, we use it to filter out circuits. Specifically, we discard 80\% of the generated circuits based on their KTA scores, which reduces the need for more computationally expensive performance evaluations.
As with classical iterative algorithms, the runtime of QuProFS depends linearly on the number of iterations, so here we analyze the cost of a single iteration. The dominant computational burden lies in the evaluation of the fidelity kernel, local effective dimension (LED), and expressibility proxies, each requiring quantum circuit executions with bounded depth \(D\) and width \(W\).

The fidelity kernel computation, leveraging the Nyström low-rank approximation with subsample size \(M \ll N\), reduces the quadratic kernel complexity from \(O(N^2)\) to \(O(N M^2)\). Specifically, the time required to compute the fidelity kernel matrix is $\tilde{O}\left(N M^{2} \frac{D W}{\epsilon_1}\right)$,
where \(\epsilon_1\) is the approximation precision controlling the accuracy of kernel estimation.

Next, the LED proxy involves estimating the Fisher information matrix eigenvalues over the parameter space with condition number \(\kappa\) and effective rank \(\mu\). Using established Fisher matrix approximations and Monte Carlo integration over an \(\epsilon\)-ball in parameter space, the runtime is $\tilde{O}\left(N \frac{D W \kappa \mu}{\epsilon_2}\right)$,
with \(\epsilon_2\) governing the precision of the Fisher matrix eigenvalue estimation.

Expressibility is estimated by sampling \(S\) pairs of PQC states and computing their fidelities. This incurs a cost of
$\tilde{O}(S D W)$ independent of \(N\) and \(M\).

Following the evaluation of these proxies, an evolutionary step is performed that sequentially prunes a portion of the gates, where the pruning amount is a hyperparameter, and appends a single layer.
Each pruning requires running the circuit once, costing \(O(C_{\text{gate}} \cdot G \cdot L)\), where \(C_{\text{gate}}\) is the cost per circuit run, \(G\) the number of pruned gates, and \(L=1\) the appended layers per step. Since these steps occur after every proxy evaluation and do not involve recompilation or retraining, their cost is $
T_{\mathrm{evo}} = O(C_{\mathrm{gate}} \cdot G),
$ which is significantly smaller than the combined proxy costs and can be safely treated as negligible asymptotically.

Combining these terms and suppressing logarithmic factors, the total runtime per iteration is
\[
\tilde{O}\left(N M^{2} \frac{D W}{\epsilon_1} + N \frac{D W \kappa \mu}{\epsilon_2} + S D W \right).
\]

This runtime reflects the dominant contribution from proxy computations, with the evolutionary step cost absorbed as a lower-order term.

\subsection{Ablation\label{annex:ablation}}
\subsubsection{Effect of metrics used}

We use different combinations of metrics to generate new circuits for 1 iteration and look and the difference in average performance of the circuits.

\begin{figure*}
\centering
\begin{tabular}{lllllll||r}
\toprule
Filter & KTA & FRO & KL & L. eff dim & CNOT & Param & B. cancer \\ 
\midrule
 \checkmark& \checkmark & \checkmark & \checkmark & \checkmark & \checkmark & \checkmark & 0.77\\ 
& \checkmark & \checkmark & \checkmark & \checkmark & \checkmark & \checkmark & 0.73\\ 
 & \checkmark &  &  &  &  &  & 0.78\\ 
 &  & \checkmark &  &  &  &  & 0.69\\ 
 &  &  & \checkmark &  &  &  & 0.73\\ 
 &  &  &  & \checkmark &  &  & $0.75 \pm 0.12$ \\ 
 & \checkmark & \checkmark &  &  &  &  & 0.75 \\ 
 & \checkmark &  & \checkmark &  &  &  & 0.76\\ 
 & \checkmark &  &  & \checkmark &  &  & 0.79\\ 
 &  & \checkmark & \checkmark &  & &  & 0.74\\ 
 &  & \checkmark &  & \checkmark &  &  & $0.71 \pm 0.11$ \\ 
 &  &  & \checkmark & \checkmark &  &  & 0.80\\ 
 & \checkmark & \checkmark & \checkmark &  &  &  & $0.73 \pm 0.12$ \\ 
 & \checkmark & \checkmark &  & \checkmark &  &  & $0.73 \pm 0.12$\\ 
 & \checkmark &  & \checkmark & \checkmark &  &  & $0.82 \pm 0.11$ \\ 
 &  & \checkmark & \checkmark & \checkmark &  &  & 0.70 \\ 
 & \checkmark & \checkmark & \checkmark & \checkmark &  &  & $0.74 \pm 0.12$\\ 
\bottomrule
\end{tabular}
\end{figure*}

We study the effect of the number of circuits evaluated per iteration on the overall performance of the algorithm. Evaluating more circuits enables a more thorough exploration of the search space, albeit at the cost of increased runtime. Depending on the effectiveness of our ranking strategy, a larger batch size may also introduce more noise that needs to be filtered.


Overall, despite increasing the number of considered circuits by more than a factor of 20, we do not observe any significant improvement across the datasets. This holds true for both kernel target alignment (KTA) and SVM test scores (without hyperparameter tuning). While we do observe some outliers in KTA, these do not translate into consistently higher SVM test scores.

We also note that many circuits yield very similar Frobenius scores, which may indicate a lack of diversity in the resulting kernels among the circuits evaluated.

\begin{figure}
    \centering
    \includegraphics[width=\linewidth]{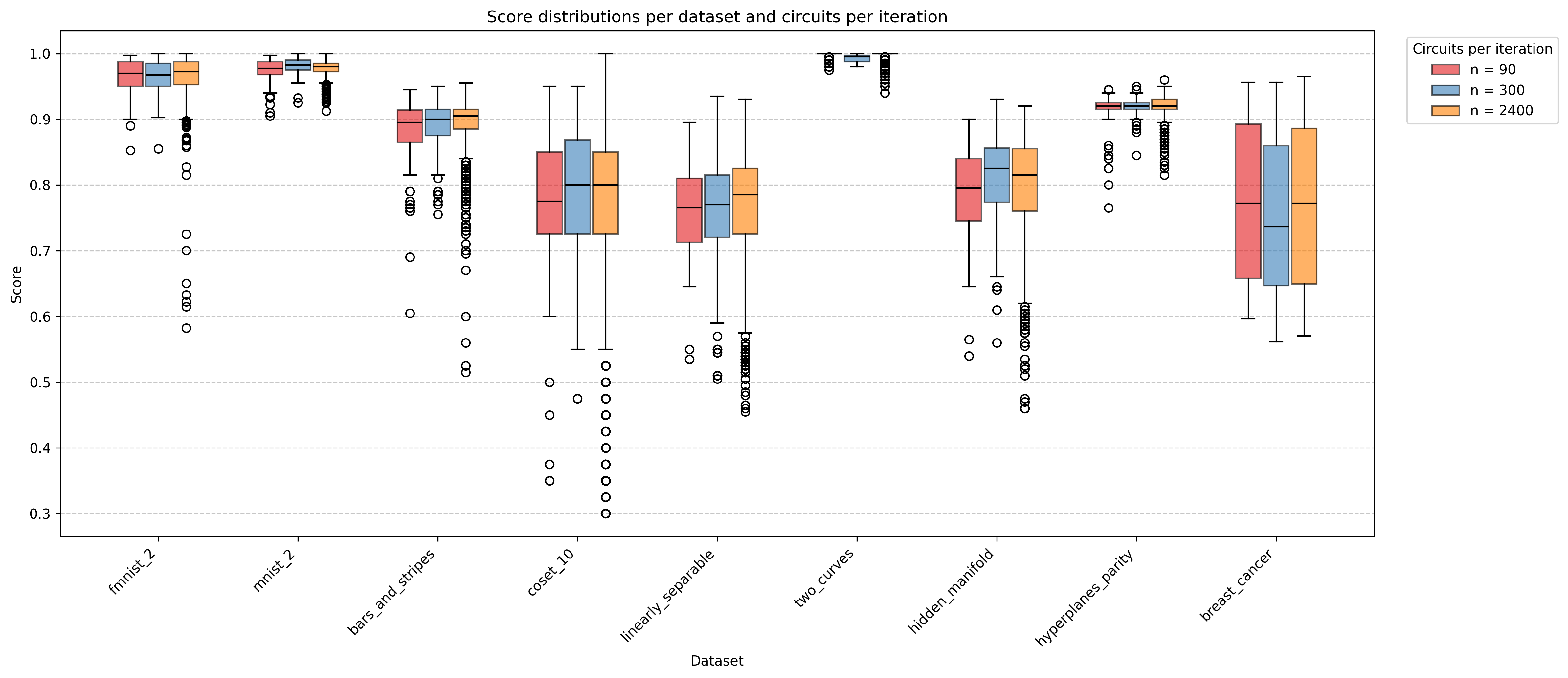}
    \caption{Box plot of the repartition of SVM test accuracy for randomly selected variational parameters across batches after one iteration for different batch sizes}
    \label{fig:test_abl}
\end{figure}

\begin{figure}
    \centering
    \includegraphics[width=\linewidth]{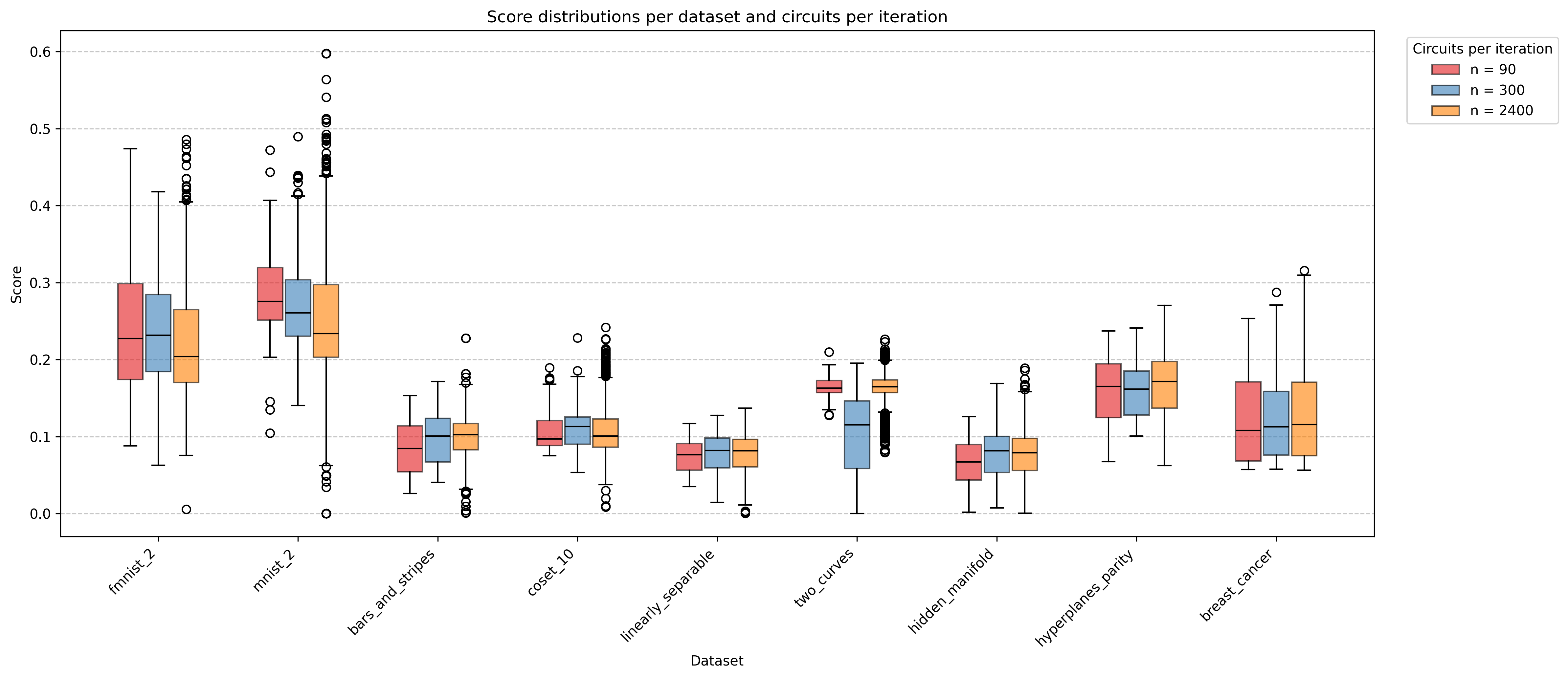}
    \caption{Box plot of the repartition of KTA across batches after one iteration for different batch sizes}
    \label{fig:kta_abl}
\end{figure}

\begin{figure}
    \centering
    \includegraphics[width=\linewidth]{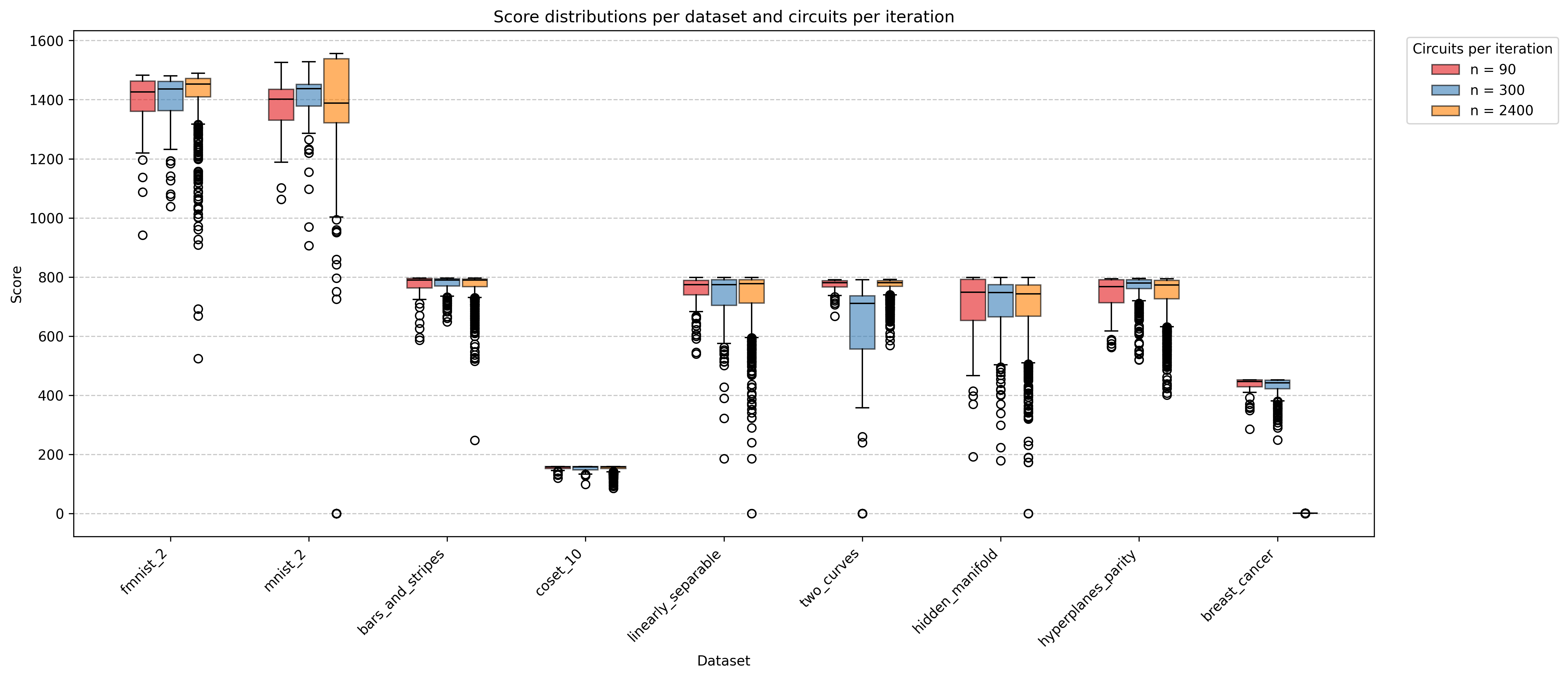}
    \caption{Box plot of the repartition of Frobenius score across batches after one iteration for different batch sizes}
    \label{fig:fro_abl}
\end{figure}

\subsection{Results~\label{annex:results}}

\begin{table*}[!ht]
\centering
\caption{QuProFS accuracy on benchmark datasets using real quantum hardware.}
\label{tab:real_hw_results}
\resizebox{\textwidth}{!}{%
\begin{tabular}{lcccccccc}
\toprule
\textbf{Dataset} & MNIST-2 & FMNIST-2 & B. \& STRIPS & H. PARITY & B. CANCER & H. MANIFOLD & L. SEPARABLE & COSET 10 \\
\midrule
\textbf{QuProFS} & \textbf{0.92} & \textbf{0.99} & \textbf{0.99} & \textbf{0.96} & \textbf{0.92} & \textbf{0.80} & \textbf{0.80} & \textbf{0.96} \\
\bottomrule
\end{tabular}
}
\end{table*}

\paragraph{Runtime Analysis}  
  
The precision parameters \(\epsilon_1, \epsilon_2\) are chosen based on desired accuracy \(\delta\) and dataset properties as detailed in Sections 3.1 and 4.1, ensuring robust proxy estimation.

\begin{theorem}[Runtime guarantee for QuProFS]
For datasets of size \(N\) with parameter dimension \(d\), under bounded circuit depth \(D\) and width \(W\), and assuming exploitable low-rank structure with Nyström subsample size \(M \ll N\), QuProFS returns near-optimal architecture proxies with high probability. The per-iteration runtime satisfies
\[
\tilde{O}\left(
N M^{2} \frac{D W}{\delta^{\alpha}} + N \frac{D W \kappa \mu}{\delta^{\beta}} + S D W
\right),
\]
where \(\delta\) controls proxy accuracy, \(\kappa\) and \(\mu\) are Fisher matrix parameters, \(S\) is expressibility samples, and \(\alpha, \beta > 0\) depend on approximation tolerances.
\end{theorem}

The runtime of QuProFS scales linearly with the number of iterations; here, we analyze the cost of a single iteration. The dominant cost arises from evaluating the fidelity kernel, local effective dimension (LED), and expressibility proxies, each requiring quantum circuit executions with bounded depth \(D\) and width \(W\).

- The fidelity kernel is approximated via the Nyström method with subsample size \(M \ll N\), reducing complexity from \(O(N^2)\) to \(\tilde{O}\bigl(N M^{2} \frac{D W}{\epsilon_1}\bigr)\), where \(\epsilon_1\) controls approximation precision.
- The LED proxy estimates Fisher information matrix eigenvalues, with runtime \(\tilde{O}\left(N \frac{D W \kappa \mu}{\epsilon_2}\right)\), where \(\kappa\) and \(\mu\) are the condition number and effective rank, and \(\epsilon_2\) controls estimation precision.
- Expressibility estimation samples \(S\) pairs of parameterized quantum circuit states, costing \(\tilde{O}(S D W)\), independent of \(N\) and \(M\).

An evolutionary step, involving pruning a portion of the gates and appending a single layer, costs \(O(C_{\mathrm{gate}} \cdot G)\), negligible compared to proxy computations.

Overall, the per-iteration runtime is
\[
\tilde{O}\left(N M^{2} \frac{D W}{\epsilon_1} + N \frac{D W \kappa \mu}{\epsilon_2} + S D W \right).
\]

\subsection{Proxy Metrics}

Our final circuit ranking leverages an ensemble of proxy metrics designed to capture complementary aspects of circuit quality. Table~\ref{tab:proxies_annex} summarizes the proxies used, while Appendix~\ref{proxies_annex} provides a detailed explanation of each. These proxies enable efficient and theoretically grounded assessments without requiring costly full circuit training or execution.

\subsubsection{Hardware robustness}
To improve circuit robustness against hardware noise, we constrain circuit design to native gates and adjacent-qubit connections, eliminating the need for transpilation and minimizing costly SWAP gates. This ensures compatibility with the device topology and reduces depth overhead.vWe further evaluate robustness using relative fidelity scores, penalizing circuits with many CNOT gates due to their high error rates. Our noise-aware cost function incorporates gate and readout errors, as well as CNOT count, enabling reliable comparison across circuits with a consistent layout. This avoids the expensive task of layout optimization, which involves subgraph isomorphism over large device connectivity graphs. 

\subsubsection{Kernel Concentration Indicator}
Quantum kernels are known to suffer from exponential concentration under certain conditions, where the kernel values converge toward a fixed constant as the number of qubits increases. This phenomenon can arise due to several factors, including the expressivity of the encoding circuit, global measurements, high entanglement, and hardware noise~\citep{thanasilp2024exponential}. This metric is used to preemptively assess whether the sampled circuits from the search space are susceptible to such concentration effects, which could limit their effectiveness in distinguishing between classes.

\subsubsection{The Overparameterization Trade-off}
We assess circuit expressivity and trainability using proxies such as the number of trainable parameters and CNOT gates. While higher expressivity can improve representational power, it also increases sensitivity to hardware noise. Although deeper, more expressive circuits are theoretically favored, they often degrade in performance on real devices due to noise accumulation.

Prior work shows that underparameterized quantum neural networks (QNNs) can get trapped in spurious local minima, impairing trainability. Overparameterization helps avoid these traps by inducing a sharp transition in the optimization landscape, as reflected in the saturation of the quantum Fisher Information Matrix (QFIM)~\citep{abbas2021power}. However, noise can reduce the effective rank of the QFIM, making even overparameterized QNNs behave as underparameterized ones~\citep{garcia2024effects}.

\subsubsection{Trainability}
Effective dimension has been introduced in the context of classical neural networks to capture the sensitivity of a model’s output with respect to perturbations in its parameter space. In the quantum setting, we adopt a similar notion to evaluate the trainability of parameterized quantum circuits (PQCs), building on the work of~\citep{abbas2021power, abbas2021effective}. Specifically, effective dimension serves as a robust proxy for the generalisation ability of a quantum neural network, leveraging the Fisher Information Matrix to characterise the geometry of the model’s parameter landscape.

The algorithm proposed in~\citep{abbas2021effective} uses Monte Carlo sampling to compute forward and backward passes of the QNN over sampled inputs and parameters—an operation that becomes the computational bottleneck. To mitigate this cost, we apply a sub-sampling strategy over the dataset, significantly reducing the runtime while maintaining fidelity in estimating the trainability metric.

\subsubsection{Expressivity/Entanglement}
The expressibility of a quantum circuit refers to its capacity to generate pure states that effectively represent the Hilbert space \citep{sim2019expressibility}. In our approach, we compute the Kullback-Leibler (KL) divergence between the distribution of states produced by sampling the parameters of a parameterized quantum circuit (PQC) and the uniform distribution of states, which is represented by Haar random states. However, since highly expressive circuits are prone to overfitting, we emphasize selecting circuits that are ranked in the middle range of expressibility.

\subsection{Proxies\label{proxies_annex}}
The main fidelity kernel costs $\mathcal{O}(n_{subsample}^2)$ where $n_{subsample}$ is the size of the subsample. Here, we look at the different proxies used in the paper in a bit more detail.

\subsubsection{Hardware robustness}
To enhance the robustness of circuits against device noise, we employ a comprehensive strategy: We design circuits that utilize only the native gates of the device and establish connections between adjacent qubits. This approach removes the need for transpilation, which generates an equivalent circuit to match the topology of a specific quantum device and makes them compatible with a given target device. Additionally, it minimizes the necessity for SWAP gates, primarily employed to connect non-adjacent qubits, which is costly to implement on real devices.

Moreover, we employ relative fidelity scores to gauge the robustness of the generated circuits in the context of hardware noise. Given that two-qubit gates like CNOT are particularly noisy, we impose a penalty on circuits that contain a high number of these gates. This is reflected in the cost function, derived from the errors associated with the gates, the readout error, and the total number of CNOT gates utilized. Since we are primarily interested in the relative ranking of the circuits, we maintain a consistent layout and calculate the cost for each circuit accordingly. This approach is necessary because finding optimal layouts involves identifying isomorphic subgraphs, which is an expensive process within the quantum processor's connectivity graph with more than 100 qubits, where each node represents a physical qubit and each edge indicates the presence of two-qubit gate support between those qubits.

While the formula employed in the mapomatic score does not cover the entire noise model of the quantum circuit, it helps assess the 
 relative performance for a set
of circuits given the layout.
This heuristic scoring function depends on the reported calibration
data for the processor. Incorporating this metric helps us evaluate how prone the circuits are to the noise in the quantum device.  
Further, a circuit will “fit” on a device if the circuit graph is subgraph-isomorphic to the device graph. Such subgraph isomorphisms are represented as layouts, mappings from the abstract circuit qubits to specific physical device qubits, as shown in Fig. 1(d)-(e). The set of all layouts for a given transpiled circuit can be efficiently identified using VF2, a well-known subgraph isomorphism algorithm [25], as well as related algorithms
The complexity of calculating the mapomatic score scales with the number of gates.
The total fidelity \( F_{\text{total}} \) of the quantum circuit is given by:
The total fidelity \( F_{\text{total}} \) of the quantum circuit is given by:
\begin{align*}
F_{\text{total}} &= \prod_{\text{gates}} \left( 1 - F_{\text{gate error}} \right) \\
&\times \prod_{\text{measurements}} \left( 1 - F_{\text{readout error}} \right) \\
&\times \prod_{\text{idle periods}} \left( 1 - F_{\text{idle error}} \right)
\end{align*}

where $F_{\text{readout error}}$ represents the readout error, $F_{\text{gate error}}$ represents the gate error for all the gates, and 
$F_{\text{readout error}}$ represents the decoherence error affecting the idle qubits.
\paragraph{Degree of Uniformity}
Quantum kernels are prone to exponential concentration under specific conditions wherein the values of the kernel concentrate around a fixed value with increasing number of qubits. This could be due to reasons such as expressivity of the encoding quantum circuit, global measurements, entanglement and noise \citep{thanasilp2024exponential}. We preemptively estimate if the circuits sampled from the search space are prone to uniformity or exponential concentration by looking at a simple metric calculated as 
\[
\|\mathcal{F} - \mathbf{1}\|_F = \sqrt{\sum_{i=1}^{n} \sum_{j=1}^{n} (f_{ij} - 1)^2}
\]
where the frobenius norm is calculated between the fidelity kernel \( \mathcal{F} \) of size \( n \times n \) and the matrix of all ones \( \mathbf{1} \) of size \( n \times n \). This score also gives an indication of how far the kernel is from uniformity. 
The complexity depends on the number of shots $d$ and the subsample size $n_{subsample}$ as $\mathcal{O}(dn_{subsample}^{2})$. 

\paragraph{Trainability}
fits and performs badly on new data samples.

For $\Theta \subset \mathbb{R}^d$ as a $d$-dimensional parameter space and $n \in \mathbb{N}, n > 1$ as the number of data samples, the effective dimension \citep{chen2024evolutionary} for a PQC  
$d_{\gamma, n}({V}_\Theta) $ with respect to $\gamma \in (0, 1]$ is given by:

In general, we should expect the value of the local effective dimension to decrease after training. This can be understood by looking back into the main goal of machine learning, which is to pick a model that is expressive enough to fit the data, but not too expressive that it overfits and performs badly on new data samples.

Certain optimizers help regularize the overfitting of a model by learning parameters, and this action of learning inherently reduces a model’s expressiveness, as measured by the local effective dimension. Following this logic, a randomly initialized parameter set will most likely produce a higher effective dimension than the final set of trained weights, because that model with that particular parameterization is "using more parameters" unnecessarily to fit the data. After training (with the implicit regularization), a trained model will not need to use so many parameters and thus have more "inactive parameters" and a lower effective dimension.

The local effective dimension of a statistical model $M_\Theta := \{ p(\cdot, \cdot; \theta) : \theta \in \Theta \}$ around $\theta^* \in \Theta$ with respect to $n \in \mathbb{N}$, $\gamma \in \left( \frac{2\pi \log n}{n}, 1 \right]$, and $\varepsilon > \frac{1}{\sqrt{n}}$ is defined as
\[
d_{n, \gamma} (M_{B_\varepsilon(\theta^*)}) = 
2 \log \left( \frac{1}{V_\varepsilon} \int_{B_\varepsilon(\theta^*)} \sqrt{\det \left( \text{Id}_d + \kappa_{n, \gamma} \overline{F}(\theta) \right)} \, d\theta \right) \frac{1}{\log \kappa_{n, \gamma}},
\]
where $B_\varepsilon(\theta^*)$ is a neighborhood of $\theta^*$ and $\kappa_{n, \gamma}$ is a function related to the model. Due to the use of Monte Carlo sampling for approximating the Fisher information matrix, the complexity of the algorithm is $\mathcal{O}(\Theta n_{subsample})$ where $\Theta$ is the number of trainable parameters and $n_{subsample}$ is the number of subsamples.

The effective dimension is given by:
\[
2 \log \left( \frac{1}{V_\Theta} \int_{\Theta} \det \left( \mathbb{I}_d + \frac{\gamma}{n} \frac{2}{\pi} \log n \hat{F}(\theta) \right) d\theta \right) \bigg/ 
\]
\[ \log \left( \frac{\gamma}{n} \frac{2}{\pi} \log n \right)
\]
where $V_\Theta := \int_\Theta d\theta \in \mathbb{R}^+$ is the volume of the parameter space and $\hat{F}(\theta) \in \mathbb{R}^{d \times d}$ is the normalized Fisher information matrix, defined as:

\[
\hat{F}_{ij}(\theta) := \frac{d}{V_\Theta} \int_\Theta \text{tr} \left( F(\theta) \right) d\theta F_{ij}(\theta)
\]
with the normalization ensuring that:
\[
\frac{1}{V_\Theta} \int_\Theta \text{tr} \left( \hat{F}(\theta) \right) d\theta = d
\]
The Fisher information matrix $F(\theta)$ is approximated by the empirical Fisher information matrix:
\[
\hat{F}_k(\theta) = \frac{1}{k} \sum_{j=1}^k \frac{\partial}{\partial \theta} \log p(x_j, y_j; \theta) \frac{\partial}{\partial \theta} \log p(x_j, y_j; \theta)^\top
\]
In this work, we consider only local effective dimension as it shows positive correlation with global effective dimension while requiring lesser time to compute.

\subsubsection{The Overparameterization Trade-off}

We consider circuit expressivity and trainability. These are estimated using proxies such as the number of trainable parameters and CNOT gates. While increasing these factors generally enhances entanglement and expressivity, it also raises the risk of noise-induced degradation. Moreover, although deeper and more expressive circuits are often favored theoretically, they may suffer in practice due to accumulated noise on real hardware.

Interestingly, prior work has shown that underparameterized quantum neural networks (QNNs) are prone to spurious local minima in their optimization landscapes, which hinder trainability. Overparameterization—by adding sufficient parameters—can eliminate these traps, leading to a sharp transition in the trainability behavior of QNNs. This "computational phase transition" is marked by the saturation of the quantum Fisher Information Matrix (QFIM), which reflects the capacity of the model~\citep{abbas2021power}. However, noise can effectively reduce a circuit's rank in the QFIM, causing even an overparameterized QNN to behave like an underparameterized one~\citep{garcia2024effects}, thus undermining its trainability.

Therefore, our evaluation framework considers these trade-offs holistically—balancing expressivity, parameter count, and hardware noise. We build upon the Mapomatic heuristic~\citep{mapomatic} to rank circuits in a hardware-aware, noise-resilient manner, ensuring deployability without sacrificing learning capability.

\paragraph{Expressivity/Entanglement}
The expressibility of a quantum circuit  
indicates its ability to generate (pure) states that are well representative of the
Hilbert space \citep{sim2019expressibility}. We follow the approach in \citep{sim2019expressibility} by computing the KL-divergence between the distribution of states obtained from
sampling the parameters of a PQC to the 
uniform distribution of states which is represented by the Haar random states.

The complexity is bounded by the number of shots $n_{shots}$ and the number of subsamples used as $\mathcal{O}(n_{subsample}n_{shots})$.

 We represent a state generated by a PQC as \( U(\theta) |0\rangle \), where the parameter \( \theta \) is randomly selected from a uniform distribution over \([0, 2\pi]\). The fidelity \( F \) between two states \( \phi_{1} \) and \( \phi_{2} \) is given by \( | \langle \phi_{1}|\phi_{2} \rangle|^{2} \). The distribution of state fidelities generated by the PQC is denoted as \( \hat{P}_{PQC}(F; \theta) \). In comparison, \( P_{Haar}(F) \) represents the distribution of state fidelities derived from \( N \)-qubit Haar random states, characterized by the analytical form \( (2^{N} - 1)(1 - F)^{2^{N} - 2} \). 

\[ Expr = \mathrm{KL}({Pr}_{\text{emp}} \parallel {Pr}_{\text{Haar}}) \]
\[ = \sum_{j} {Pr}_{\text{emp}}(j) \log \left( \frac{{Pr}_{\text{emp}}(j)}{{Pr}_{\text{Haar}}(j)} \right)
\]

The expressibility is then estimated as the Kullback-Leibler divergence between the empirically observed probability distribution and the probability distribution expected under a Haar-random unitary. Lower $Expr$ value indicates that a PQC can produce a broader variety of states within Hilbert space, reflecting enhanced expressibility.
\paragraph{Runtime Analysis}

We consider the absolute runtimes per architecture (i.e circuit). For the SOTA models considered such as Elivagar, QSEA and QNAS, we consider the runtime with and without the training of the circuits. This is due to the difference in the training process for QNN vs quantum kernels. Due to the differences in the parallelization and hardware specifications in SOTA methods, we note the difference in runtimes from that cited in the \citep{anagolum2024elivagar, chen2023quantumsea, wang2022quantumnas}. Our experiments were conducted on a system with the following hardware specifications: Red Hat Enterprise Linux OS, an Intel(R) Xeon(R) Platinum 8360Y CPU @ 2.40GHz, eight NVIDIA A100-SXM4-40GB GPUs, and 500GB of memory.

\paragraph{Data-set compatibility\label{KTA_annex}}
 Kernel Target Alignment (KTA) measures the similarity between two kernels or the degree of agreement between a kernel and a dataset \citep{cristianini2001kernel}. Recently, KTA has popularity as replacement for the loss function based on classification accuracy for QSVM to improve the evaluation speed as evaluating the accuracy is expensive \citep{hubregtsen2022training}.In this work, we use this measure over a sub-sampled dataset to measure the compatibility between the task and the chosen
kernel. For a given labeled dataset $\mathcal{D} = \{(x^1,y^1),\ldots,(x^m,y^m)\}$ with $m$ samples and class labels $\forall i, y_i \in {-1, 1}$,  KTA can be calculated as
$$ KTA = \frac{{\langle K, O \rangle}_{F}}{\sqrt{\langle K, K \rangle_{F}\langle O, O \rangle}_F} $$

where the kernel function $\kappa(x_i, x_j)$ is used to calculate $K$ as 
$$ K_{ij} = k(x_i, x_j)$$ and with the oracle matrix $O_{n \times n}$ defined using the training labels as $$ O_{ij} = y_{i}y_{j}^{T}$$

The complexity of calculating KTA depends on the number of subsamples $n_{subsample}$ and involves calculating the calculating of inner product giving it $\mathcal{O}(n_{subsample}^{2})$

For each dataset, we standardize input features using \texttt{StandardScaler}. Dimensionality reduction using PCA \citet{mackiewicz1993principal} is applied when appropriate to ensure compatibility with the available number of qubits (e.g., 8 dimensions for Breast Cancer). Datasets are split into 80\% training and 20\% testing. For binary image-based datasets such as Bars and Stripes, we use 4$\times$4 patterns (16 qubits). Coset datasets use 14–20-dimensional inputs mapped directly to qubits. We deploy quantum circuits on real quantum hardware from IBM Quantum, including \texttt{ibmq\_brisbane}, \texttt{ibmq\_strasbourg}, \texttt{ibm\_kyiv}, and \texttt{ibm\_sherbrooke}, with backend selection guided by device availability and qubit connectivity. Due to hardware constraints, we evaluate the top-ranked architectures on 8 representative datasets.

We randomly select approximately 50\% of the circuits from the current pool in each refinement cycle. For each selected circuit, we apply gate pruning with a probability of 40\%, meaning roughly 40\% of the gates are removed. This pruning rate is a tunable hyperparameter that we found effective in balancing exploration and performance within our experimental context.

\end{document}